\documentclass[conference]{IEEEtran}
\ifCLASSINFOpdf
\else
\fi
\usepackage{fancyhdr}

\usepackage{algorithm}
\usepackage{xspace}
\usepackage[font=small,labelfont=bf]{caption}
\usepackage{amsmath}
\usepackage{mathtools}
\usepackage{xcolor}
\usepackage{amsfonts}
\usepackage{amssymb}
\usepackage{pifont}
\usepackage{color}

\usepackage{algpseudocode}

\usepackage{textgreek}
\usepackage{subfig}
\usepackage{textcomp}
\usepackage[]{siunitx}

\makeatletter
\algrenewcommand\ALG@beginalgorithmic{\footnotesize}
\makeatother

\usepackage{fancyhdr}
\usepackage[normalem]{ulem}
\usepackage[hyphens]{url}
\usepackage[sort,nocompress]{cite}
\usepackage[final]{microtype}
\usepackage{flushend}
\usepackage{makecell}

\usepackage[bookmarks=true,breaklinks=true,letterpaper=true,colorlinks,linkcolor=black,citecolor=blue,urlcolor=black]{hyperref}

\algnewcommand{\LineComment}[1]{\State \(\triangleright\) #1}

\xspaceaddexceptions{]\}}
\newcommand{\ignore}[1]{}

\newcommand{\old}[1]{}
\newcommand{\fig}[1]{Figure~\ref{#1}}
\newcommand{\sect}[1]{Section~\ref{#1}}
\newcommand{\tab}[1]{Table~\ref{#1}}
\newcommand{\algo}[1]{Algorithm~\ref{#1}}
\newcommand{\eqn}[1]{Equation~\ref{#1}}

\newcommand{\drain}[1]{\texttt{DRAIN}\xspace}

\newcommand{\cpuonly}[0]{\texttt{CPU-only}\xspace}

\newcommand{\cpugpu}[0]{\texttt{CPU-GPU}\xspace}
\newcommand{\nmpgpu}[0]{\texttt{NMP-GPU}\xspace}

\newcommand{\baseline}[0]{\texttt{Baseline(CPU)}\xspace}
\newcommand{\ourscc}[0]{\texttt{Ours(CPU)}\xspace}
\newcommand{\oursmc}[0]{\texttt{Ours(NMP)}\xspace}

\newcommand{\dlrm}[1]{RM{#1}\xspace}

\newcommand{\tcasting}[0]{Tensor Casting\xspace}
\newcommand{\tcasted}[0]{\texttt{T.Casted}\xspace}

\newcommand\blfootnote[1]{%
\begingroup
\renewcommand\thefootnote{}\footnote{#1}%
\addtocounter{footnote}{-1}%
\endgroup
}

\renewcommand{\baselinestretch}{.999}
\title{\huge Tensor Casting: Co-Designing Algorithm-Architecture \\for Personalized Recommendation Training
} 

\begin{document}

\author{

\IEEEauthorblockN{
Youngeun Kwon\hspace{2.3em}Yunjae Lee\hspace{2.6em}Minsoo Rhu\hspace{1em}}
\IEEEauthorblockA{
School of Electrical Engineering\\
KAIST\\
\texttt{\{yekwon, yunjae408, mrhu\}@kaist.ac.kr}\\
}
}


\maketitle
\pagestyle{plain}

\begin{abstract}

Personalized recommendations 
are one of the most widely deployed machine learning (ML) workload
serviced from cloud datacenters.
As such, architectural solutions for high-performance 
	recommendation inference have recently been the target of 
				several prior literatures. Unfortunately, little have been explored 
				and understood regarding the training side of this emerging ML workload.
				In this paper, we first perform a detailed workload characterization study
				on training recommendations, root-causing sparse embedding layer training
				as one of the most significant performance bottlenecks. We then 
				propose our algorithm-architecture co-design called
				Tensor Casting, which enables the development of a generic
				accelerator architecture for tensor gather-scatter that  
encompasses all the key primitives of training embedding layers.
				When prototyped on a real CPU-GPU system, Tensor Casting provides 
				$1.9-21\times$ improvements in training throughput compared to state-of-the-art approaches.

	\end{abstract}

\IEEEpeerreviewmaketitle


\blfootnote{
This is the author preprint version of the work.
}

\section {Introduction}
\label{sect:intro}

The enormous compute 
	power machine learning (ML) accelerators (also known as \emph{tensor
				processing units}, TPUs) deliver allows 
		practitioners to develop sophisticated and powerful deep neural network
		(DNN) algorithms.  Such trend has spurred numerous academic and industrial
		proposals on designing energy-efficient TPUs for both training and
		inference.
	Interestingly, while it is challenging to define a \emph{generic} TPU
	architecture that encompasses numerous design points for ML
	inference (i.e., they are typically optimized for a \emph{diverse} set of
			compute primitives that represent a specific application domain, for
			instance, convolutions for computer
			vision~\cite{dadiannao,eyeriss_isca,eyeriss,scnn,cnvlutin,stripes,bitpragmatic,bittactical,neurocube,whatmough:2017:isscc,minerva,snapea,bitfusion,dnnweaver,maeri,maestro}), TPUs for training have more or
	less settled on a design optimized for a \emph{single} key primitive:
	general-purpose matrix multiplication (GEMM). Such trend is notably
	represented by major vendors in the ML training market optimizing their TPU
	for high-performance GEMM operations, such as NVIDIA's Tensor
	Core~\cite{volta_v100}, Google's TPU~\cite{tpu2}, Habana's Gaudi~\cite{gaudi},
	Graphcore's IPU~\cite{ipu}, and Cerebra's CS-1~\cite{cerebras}.   The reason why
	TPUs for training are optimized for this single compute primitive is because
	of the \emph{versatility} of GEMM operators. Concretely, the backpropagation
	based training algorithm~\cite{sgd} requires the derivation of input and weight
	gradients, both of which are able to be \emph{casted} as a GEMM operation for
	popular DNN layers, such as convolutional, fully-connected, and recurrent
	layers.  As these GEMM-compatible operations cover a significant portion of
	the training time for conventional DNNs, the design cost of a TPU
	can be amortized by optimizing its microarchitecture for this all-round compute
	primitive.  As such, we have witnessed great success of these GEMM-optimized
	TPUs for the purpose of training DNNs for computer vision, speech
	recognition, natural language processing, etc. 

	Despite such success however, emerging ML applications are now evolving into
	having complex DNN topologies with algorithmic components that cannot be
	singlehandedly accelerated with these GEMM-optimized TPUs.  For instance, ML
	algorithms for natural language processing~\cite{bert}, memory-augmented neural
	networks~\cite{mann}, or recommender systems~\cite{ncf,facebook_dlrm,youtube_recsys} employ
	``sparse'' \emph{embedding layers} whose compute and memory access
	characteristics significantly differ compared to conventional ``dense'' DNN
	layers.  In particular, personalized recommendation models for consumer facing products (e.g., e-commerce, Ads) stand
	out with their high memory capacity and bandwidth
	demands~\cite{tensordimm,recnmp,deeprecsys}, rendering conventional dense-optimized TPUs
	suboptimal in handling the training process of recommendations.  Because of
	the \emph{embedding tables} they utilize, the memory capacity demands of
	state-of-the-art recommendation models are in the order of several tens
	to thousands of GBs, far exceeding the memory capabilities of TPU devices.
	Consequently, existing solutions for recommendations almost exclusively
	utilize the host CPU memory~\cite{aibox,fb:zion} as a placeholder to store these
	embedding tables to overcome the limited memory size of
	TPUs. Facebook's Zion large-memory unified training
		platform~\cite{fb:zion} for instance employs a large 
			pooled CPU memory for storing memory-capacity limited embedding
			tables, with Baidu's AIBox~\cite{aibox} targeting recommendation models requiring
	several TBs of storage.  Now, an interesting property of training recommendation models
			is that these \emph{memory-capacity limited embedding tables are the
				model parameters that are subject for training}.  Therefore, training
				recommendation models takes a \emph{hybrid} approach where the sparse
				embedding layers are trained by the CPU while the dense DNN layers
				are trained using the TPU. As we detail in
				\sect{sect:time_breakdown}, such \emph{CPU-centric} training of embeddings (i.e.,
						the key compute primitives of both forward and backward propagation
						of embedding layers are executed by the CPU) causes serious
				system-level bottlenecks because of the low compute and memory
				throughput of CPUs, with the TPU contributing to less than 
				$6\%$ of the training time for several embedding intensive models.
				This raises questions on the cost-effectiveness of dense TPUs for
				training recommendations, as the majority of training time can be spent
				``outside'' the TPU accelerator, for instance, the CPUs.

	Unfortunately, little has been studied and understood in the computer
	systems community regarding the algorithmic properties of training DNN-based
	recommendation models let alone the accelerator system architecture that
	enables it.  As such, an important motivation and {\bf key contribution}
	of our study is a detailed
	analysis and characterization of the training process of DNN-based
	recommendations. Several recent work~\cite{tensordimm,recnmp}
	identified and addressed the performance bottlenecks caused by the
	\emph{tensor gather-and-reduce} operation, a key primitive of the
	``inference'' pass of recommendations. A unique contribution of our
	characterization is the revelation of a new system-level bottleneck in
	``training'' recommendation models, the \emph{gradient expand-and-coalesce}
	operator, which is a highly throughput-hungry compute primitive undertaken by
	the CPU during the backpropagation stage of embedding layers.

	To address the system-level bottlenecks of baseline CPU-centric
	systems, we propose an algorithm-architecture co-design for high-throughput
	recommendation training. The proposed co-design is based on our novel
	\emph{Tensor Casting} algorithm that transforms the backpropagation's
	gradient expand-and-coalesce primitive into a tensor gather-reduce
	operation\footnote{As gradients are structured as tensors, we
		interchangeably refer to \emph{casted} operation as both tensor gather-reduce and
			gradient gather-reduce in this paper.
	} which is, as noted above, the key compute primitive of the forward
	propagation of embedding layers (we detail why \tcasting transforms
			it into a gather-reduce later in this section). The {\bf key innovation} of our
	\tcasting is twofold. First, \tcasting algorithmically
	guarantees that the memory intensity of the gradient expand-and-coalesce
	operation is reduced by $2\times$ when it is casted to a tensor
	gather-reduce operator, significantly reducing the latency in conducting
	backpropagation of embedding layers.  Additionally, \tcasting utilizes
	the current software stack as-is so another key advantage of our proposal is
	that it can be implemented purely in software. We implement \tcasting 
	using PyTorch~\cite{torch} and CUDA, quantitatively demonstrating that our
	proposal improves the end-to-end training performance of current CPU-centric systems
	by $1.2$$-$$2.8\times$. 

While the performance improvement offered by our \tcasting algorithm is 
impressive, we observe that the baseline CPU-centric system does not fully reap out
the potential of our proposal.  To maximally utilize the
opportunities inherent with our algorithm-architecture co-design, the second
key innovation of our work is the development of a \emph{memory-centric}
training system for recommendations. Recall that the tensor gather-reduce
and the gradient expand-and-coalesce operators are
the primary computational building block of the forward and backward propagation of 
embedding layers, respectively. When \tcasting is used to transform the
gradient expand-and-coalesce into a tensor gather-reduce operator, ``both'' forward
and backward propagation are now able to be executed using a \emph{common}
compute primitive: tensor gather-reduce.
This opens
	up a unique opportunity to design a \emph{generic} accelerator
	microarchitecture that covers both the forward and backpropagation of
	embedding layers. 

	As such, our development of \tcasting to permute gradient
	expand-and-coalesce into tensor gather-reduce is intentional: similar to
	how the GEMM-optimized TPU architecture has been instrumental in the success
	of training \emph{dense} and \emph{compute-intensive}	DNNs, we argue
	that emerging ML workloads using embeddings deserve a targeted acceleration
	treatment tailored to the \emph{sparse} and \emph{memory-intensive} dataflow
	of training embedding layers.  Building on top of a recently proposed near-memory
	processing (NMP) solution for recommendation
	inference~\cite{tensordimm,recnmp}, the novelty of our memory-centric approach
	is the utilization of the \emph{expressive power of the tensor
		gather-reduce} operator to develop a generic, \emph{sparse}-optimized
		NMP accelerator for training embedding layers.  We demonstrate that our
		NMP accelerator for tensor gather-reduce can accelerate 
		the most time-consuming primitives of recommendation training.  The effectiveness of
		our memory-centric system is demonstrated as a proof-of-concept 
		prototype on a high-end GPU system, achieving an additional
		$1.5$$-$$16\times$ training time reduction than our software-only
		CPU-centric system optimized with \tcasting ($1.9-21\times$ speedup when
				compared against the baseline CPU-centric systems).

\section{Background and Related Work}
\label{sect:background}

\begin{figure}[t!] \centering
	\includegraphics[width=0.43\textwidth]{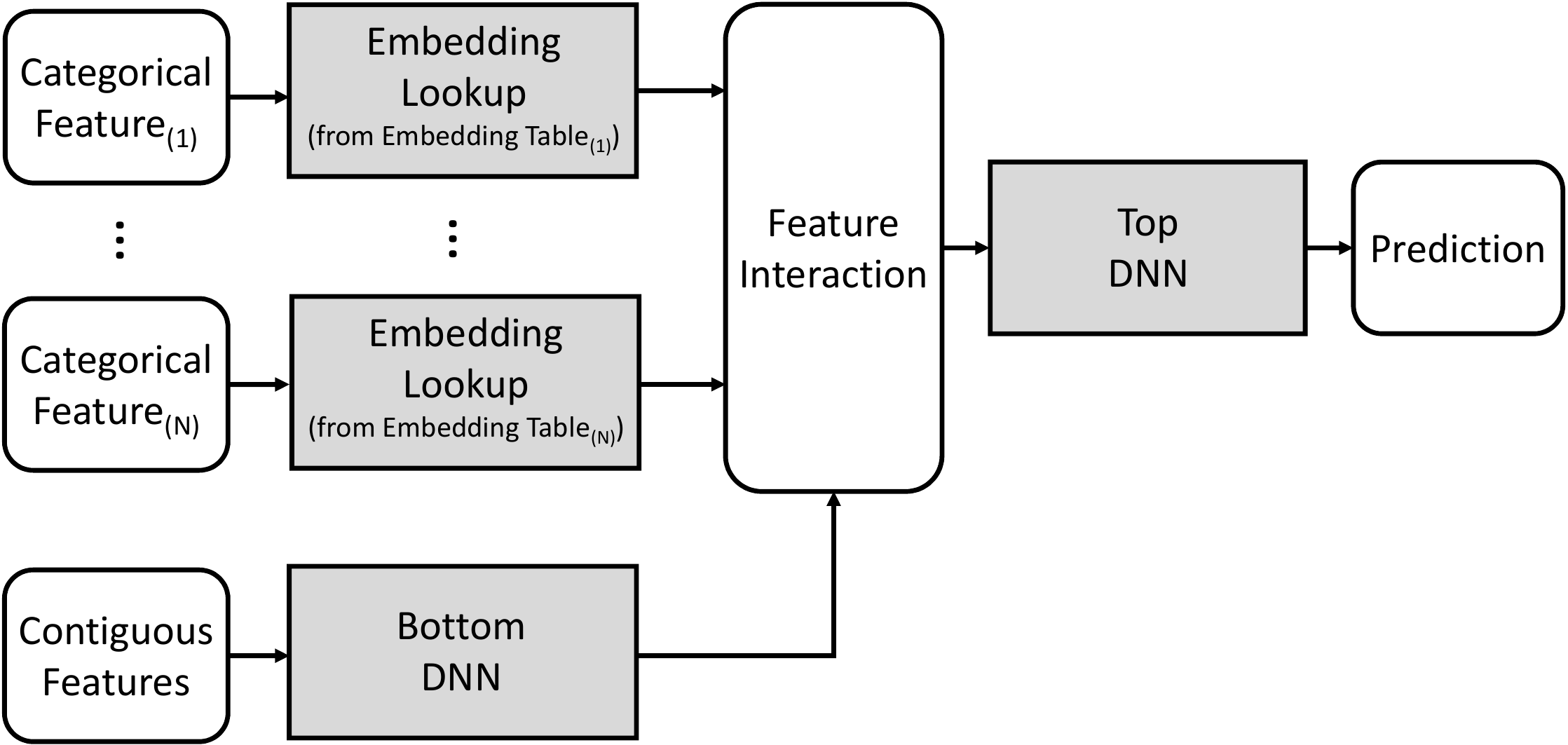}
\vspace{0.5em}
\caption{
Topological structure of a DNN-based recommendation model.
}
\vspace{-1.2em}
\label{fig:emb_layer_fwd}
\end{figure}

\subsection{Embedding Layers in Recommendations}
\label{sect:recsys_101}

The ability to ``learn'' a semantically meaningful 
	representation of a target feature is one of the biggest
		strengths of ML.  Rather than using one-hot encoded vectors for
		representing categorical features, recent work~\cite{bert,mann,facebook_dlrm} has shown that
		``\emph{embeddings}'' can be used for learning the semantic
		representation of categorical data.
For example, different movies 
		serviced by online video streaming service (e.g., YouTube, Netflix) are projected from a discrete movie ID  (i.e., each movie
		is assigned with a unique ID) to a corresponding continuous vector representation using the \emph{embedding layers}.
An embedding \emph{table} is utilized to map
		a discrete, categorical feature into its corresponding vector representation, which
		is stored contiguously within the memory
		address space as a single dimensional array. 
	Recommendation systems are formulated as a problem of predicting the
		likelihood of a certain event (e.g., the probability of a YouTube user
				watching a recommended video clip).  Figure~{\ref{fig:emb_layer_fwd}}
	provides a high-level overview of a DNN-based recommendation system, which
		consists of a frontend embedding layer and a backend DNN layer.  As there
		exists multiple categorical features (e.g., user ID, item ID, $\ldots$)
		that are helpful in capturing various semantic representations, multiple
		embedding tables are utilized.

		\begin{figure}[t!] \centering
\subfloat[]
{
	\includegraphics[width=0.485\textwidth]{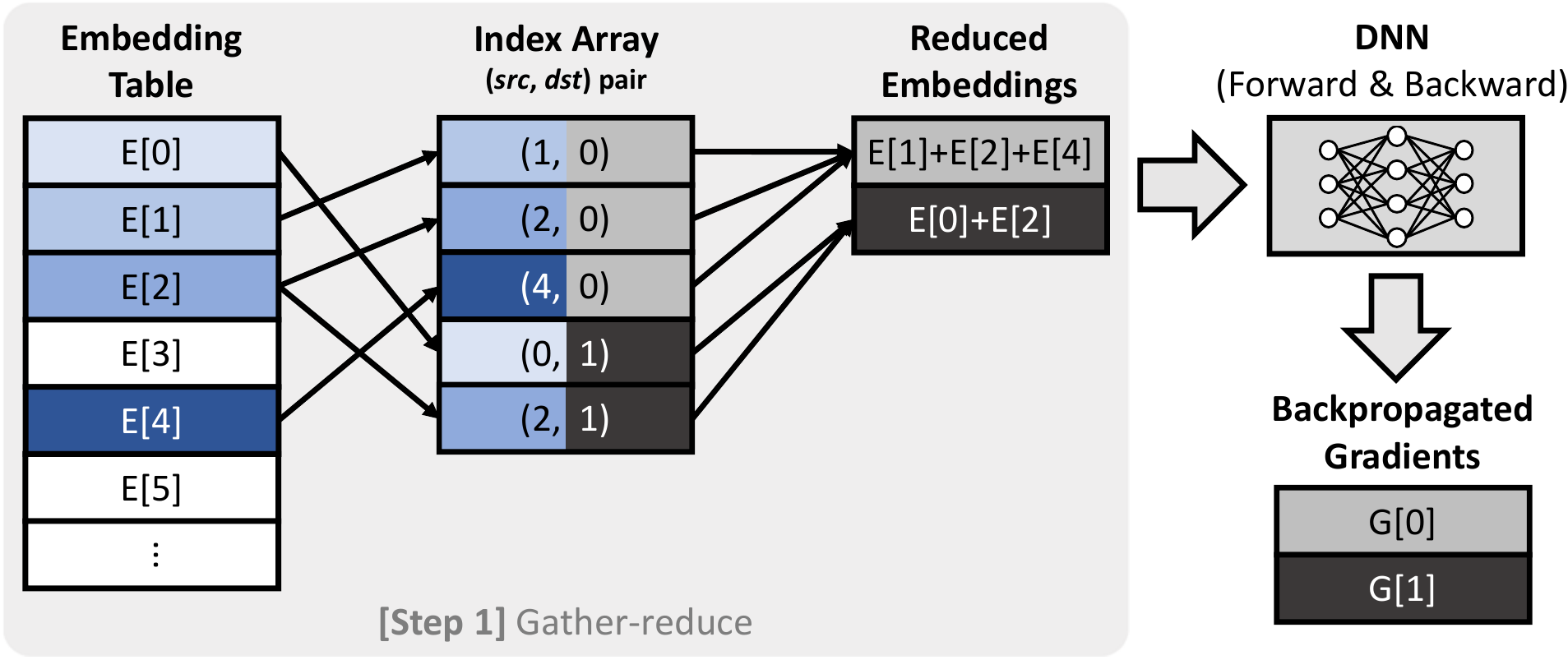}
}
\vspace{-0.8em}
\subfloat[]
{
\includegraphics[width=0.485\textwidth]{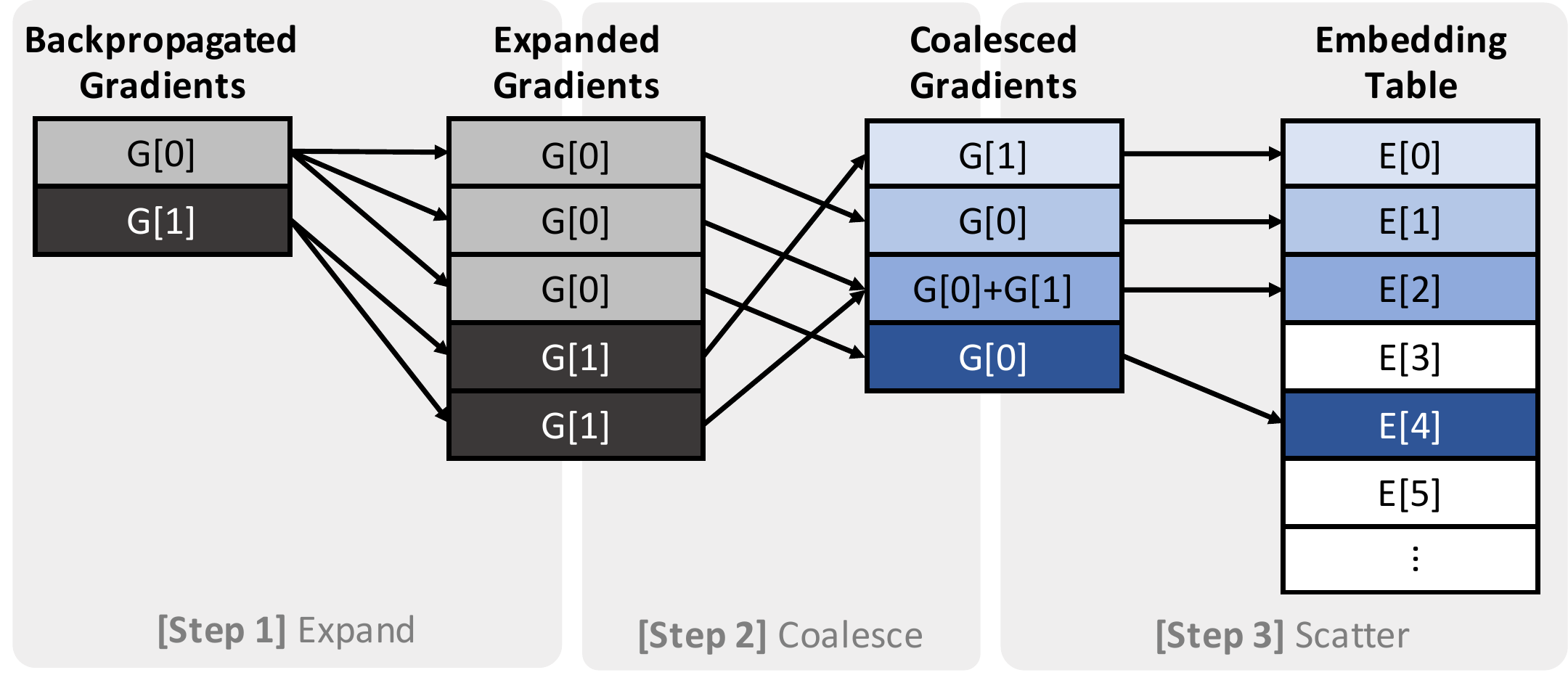}
}
\vspace{0em}
\caption{
High-level overview of (a) forward and (b) backpropagation of embedding
	layers. Example assumes a batch size of $2$, each batch input gathering $3$
	and $2$ embedding vectors, respectively. Although the embedding gather-reduce
	example in (a) appears as if it is a two separate steps (i.e., gather first
			using $src$, then reduce at $dst$), they are actually implemented as
	a \emph{fused} kernel to save memory bandwidth. Concretely, the gather-reduce
	operator utilizes the \emph{in-place} operative nature of reductions to conduct
	the gathered embedding vectors' reduction ``on-the-fly'' inside
	the on-chip registers.
}
\vspace{-1.3em}
\label{fig:recsys_training}
\end{figure}

\subsection{Training DNN-based Recommendations}
\label{sect:layer_type}


{\bf Forward propagation.} \fig{fig:recsys_training}(a) illustrates 
the forward propagation step of training embedding layers in
DNN-based recommendation models~\cite{facebook_dlrm}. An embedding table
contains millions to billions of embedding vectors, each of which is typically
sized at several hundreds of bytes, rendering a single embedding table to amount to 
several tens of GBs. As multiple embedding tables are utilized for a
recommendation model, the aggregate size of embedding tables can reach
several tens to thousands of GBs~\cite{facebook_dlrm,aibox,tensordimm}.  The
embedding \emph{gather} operation utilizes an array of index IDs 
(provided as part of the training dataset)
to lookup
multiple rows from the tens of GBs of embedding table.  While the embedding
table itself is stored in memory in a dense manner, gathering embedding vectors
exhibits a highly \emph{sparse} and \emph{irregular} memory access pattern with
low locality.  The gathered embedding vectors are then combined with each other
through an element-wise operation (e.g., addition, multiplication, etc.), which
is equivalent to a \emph{reduction}.  As illustrated in the example in
\fig{fig:recsys_training}(a), the index array for gather-reduce consists of a
($src$, $dst$) pair which is utilized to determine which element to lookup
(using $src$ ID) from the table and where to store the reduced embedding vector
(using $dst$ ID).  The reduced tensor is then fed into backend DNN layers which
is typically constructed using MLPs, followed by a softmax layer that derives
the click-through rate (CTR), e.g., the likelihood of an end-user clicking an
Ad banner. The backpropagation step of DNN layers then generates a set of
gradient vectors to backpropagate into embedding layers. The number of
gradients to backpropagate is equivalent to the number of reduced embeddings
during forward propagation, which is $2$ (i.e., \texttt{G[0]} and \texttt{G[1]}) 
	in the given example.

\begin{algorithm}[t!]
	\small
	\caption{Gradient Coalescing}
	\label{algo:coalescing}
	\begin{algorithmic}[1]
	\Procedure{Coalescing}{$src$, $exp\_grad$}
		\LineComment{Step-A: {\bf Sort} $src$ array to determine coalescable indices}
		\State $n$ $\leftarrow$ length($src$)
		\State $sorted\_pos$ $\leftarrow$ \Call{ArgSort}{$src$}
		\State $sorted\_src$ $\leftarrow$ \Call{Sort}{$src$}
		\LineComment{Step-B: {\bf Accumulate} coalescable gradients}
		\State ($i$, $prev$) $\leftarrow$ (-1, -1)
		\For {$j$ $\leftarrow$ 0 to $n$}
			\State $pos$ $\leftarrow$ $sorted\_pos$[$j$]
			\State $curr$ $\leftarrow$ $sorted\_src$[$j$]
			\If {$curr$ != $prev$}
				\State $i$ $\leftarrow$ $i + 1$
				\State $coal\_grad$[$i$] $\leftarrow$ $expanded\_grad$[$pos$]
			\Else
				\State $coal\_grad$[$i$] $\leftarrow$ $coal\_grad$[$i$] + $exp\_grad$[$pos$]
			\EndIf
		\EndFor
		\State \Return $coal\_grad$
	\EndProcedure
	\end{algorithmic}
\end{algorithm}

{\bf Backpropagation.} Training involves updating the model parameters of the
DNN, which is done using stochastic gradient descent (SGD)~\cite{sgd}.  Unlike
conventional dense DNNs, an interesting property of ML applications
utilizing embedding layers is that the contents inside the embedding tables
(i.e., the embedding vectors) are crucial part of the model that are being
trained.  Consequently, the embeddings that have been gathered and
subsequently reduced during forward propagation (e.g., $src$ index ID $1$, $2$,
		and $4$ for the first batch and ID $0$ and $2$ for the second batch in
		\fig{fig:recsys_training}) must be updated (i.e., trained) during
backpropagation.  \fig{fig:recsys_training}(b) shows how the two gradients,
	backpropagated from the backend DNN layers, are used to update multiple rows
	within the embedding table (i.e., \texttt{E[0,1,2,4]}, those that have been
			looked up during forward propagation), which is effectively a gradient
	\emph{scatter} operation back to those $src$ indices used for embedding
	gathers (i.e., the gradient scatter operator is the \emph{dual} operation of
			tensor gather for backpropagation).  Note that, a given embedding vector
	can be read out multiple times for reduction operations across
	\emph{different} batches (e.g., embeddings for $src$ ID$=$$2$ is gathered
			twice for the first and second batch). Under such circumstances, the
	corresponding embedding vector has \emph{multiple} gradient values to account
	for during model updates. That is, the multiple gradients must be
	accumulated (or \emph{coalesced}) first before being utilized for updating
	the model parameter, i.e., \texttt{G[0]}+\texttt{G[1]} is used for updating
	the embedding vector \texttt{E[2]}.  A key reason why gradients are not
	instead \emph{sequentially} updating the model directly (which obviates the
			need for gradient coalescing) is because ML frameworks are designed to
	support a variety of optimization algorithms (e.g., RMSprop, Adagrad, momentum, $\ldots$), which require the (potentially multiple) gradients
	for updating a given model parameter at the $i^{th}$ iteration
	(\texttt{W$_{i}$}) to first be accumulated into a single value
	(\texttt{G$_{i}$}), as exemplified through the optimization functions of
	RMSprop (\eqn{eqn:rmsprop}) and Adagrad (\eqn{eqn:adagrad}) model updates.

\vspace{-0.5em}
	\begin{equation}
	\label{eqn:rmsprop}
		A_{i}={\gamma}A_{i-1}+(1-\gamma)G_{i}^2, \hspace{1em} W_{i}=W_{i-1}-lr\times\frac{G_{i}}{\sqrt{\epsilon+A_{i}}}
	\end{equation}
	\vspace{-2em}

	\begin{equation}
	\label{eqn:adagrad}
		A_{i}=A_{i-1}+{G_{i}}^2, \hspace{1em}  W_{i}=W_{i-1}-lr\times\frac{G_{i}}{\sqrt{\epsilon+A_{i}}}
	\end{equation}

As illustrated in \fig{fig:recsys_training}(b), the gradient vector for batch
$0$ and batch $1$ are first each \emph{expanded} to $3$ and $2$ vectors
respectively (i.e., the gradient expand operator is the \emph{dual} operation
		of tensor reduce for backpropagation), equivalent to the number of gathers
conducted during forward propagation. The expanded vectors (i.e., tensors) are
then \emph{coalesced} into a single vector by examining whether the two tensors
have any overlapping sparse indices (e.g., $src$ ID$=$$2$ is used for both
		batch $0$ and $1$) used for constructing the gather-and-reduced vectors
during forward propagation. \algo{algo:coalescing} details the two-step 
	procedure of how the expanded gradients are coalesced. First, the
		$src$ part of the index array of \fig{fig:recsys_training}(a) is \emph{sorted} (line $2-5$)\footnote{The \texttt{ArgSort} in line $4$ is functionally identical
			to Numpy's \texttt{argsort()}, which returns the
				indices that would sort the input array.}
		to have the coalescable indices appear as consecutive elements after
		sorting (i.e.,
						\texttt{[1,2,4,0,2]}$\rightarrow$\texttt{[0,1,{\bf 2},{\bf 2},4]},
						line $5$). Then, the sorted indices are utilized for \emph{accumulating}
				the gradient values sharing a common $src$ ID, deriving the 
				coalesced gradients (line $6$$-$$17$). In this paper, we collectively refer to
				the aforementioned procedure as 
				\emph{gradient expand-coalesce} operation\footnote{Both
					TensorFlow~\cite{tensorflow} and PyTorch~\cite{torch} employ such
						gradient expand-coalesce based approach to generate the accumulated
						gradient values.}, which generates the coalesced gradients
						(\texttt{G$_{i}$}) to be used by the optimization function for the
						\emph{scatter} operation to update the models inside the embedding
						table.

\subsection{System Architectures for Training}
\label{sect:sysarch_recsys}

\begin{figure}[t!] \centering
	\includegraphics[width=0.38\textwidth]{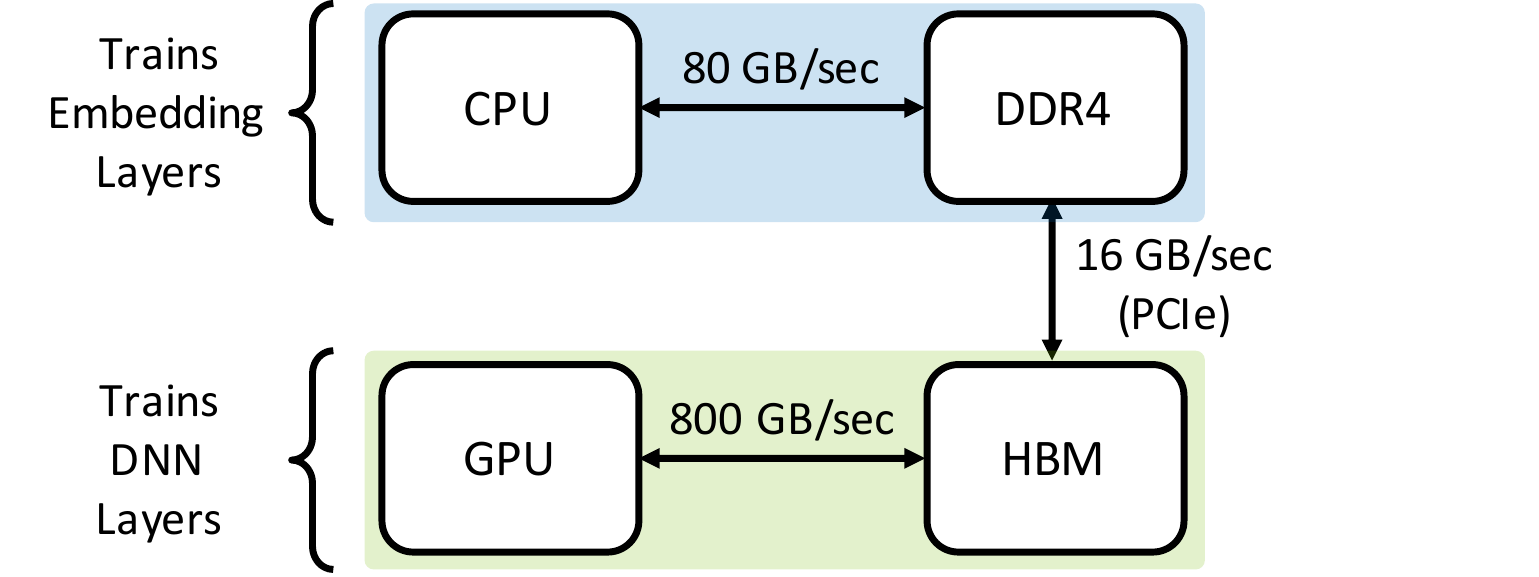}
\vspace{-0.1em}
\caption{
System architecture for training recommendation models.
}
\vspace{-1.3em}
\label{fig:sysarch}
\end{figure}

This paper assumes GPUs as the baseline TPU architecture as
it is currently the most popular platform for training purposes. Nonetheless,
	 our proposal is equally applicable for other GEMM-optimized TPUs.
State-of-the-art TPUs
for training~\cite{volta_v100,tpu2,gaudi,ipu,nnp_t} typically 
utilize bandwidth-optimized 3D stacked DRAMs such as
HBM~\cite{hbm}.  Stacked DRAMs however are capacity limited, only
available with several tens of GBs. Consequently, existing
solutions for training recommendations almost exclusively utilize either the
host CPU memory~\cite{deeprecsys,aibox,fb:zion} or a disaggregated memory
pool~\cite{tensordimm,buddy_compression} populated with capacity-optimized
DRAM (e.g., DDR4) as a separate placeholder to store the
memory-hungry embedding tables.  
Two possible system solutions for training recommendations are as follows. A
\emph{CPU-only} solution (\cpuonly) conducts training using
the CPU without GPUs~\cite{dlrm:arch}. A
\emph{CPU-GPU} based approach (\cpugpu) utilizes the CPU for
executing the embedding layers while the DNN is handled by
the GPU~\cite{aibox,nvidia:blog,tensordimm}. A common property of both
solutions is that the key primitives of training embedding layers (i.e., tensor
		gather-reduce, gradient expand-coalesce, and gradient scatter, detailed
		in \fig{fig:recsys_training}) are all conducted over the CPU. We henceforth
collectively refer to both \cpuonly and \cpugpu as \emph{CPU-centric} training
systems as the embedding layers are trained using the CPU (\fig{fig:sysarch}).

\subsection{Related Work}
\label{sect:related}

There have been several prior literature within the computer system community
focusing on the
computationally intensive CNNs, RNNs, and MLPs, which exhibit a dense
and highly regular computational property.  Because of its highly
deterministic dataflow, these dense DNN algorithms are amenable for accelerated
computation using custom-designed architectures for both training and
inference~\cite{eyeriss,whatmough:2017:hotchips,park:2018:olaware,diannao,dadiannao,cambricon,song:2015:eie,scnn,cnvlutin,fpga:dense1,
	fpga:dense2, fpga:dense3,
	fpga:dense4,wu:2019:fb_edge,rhu:2016:vdnn,mcdla,mcdla:cal,kwon:2019:disagg,rhu:2018:cdma,neummu,choi:2020:prema,jang:2019:mnnfast}.
Unlike these prior art, DNN-based recommendation models are a new type of ML workload that has only
recently been the target of study in the academic literature. 
Recent work provides a comprehensive
characterization study on \cpuonly~\cite{dlrm:arch,deeprecsys,centaur:hwang} and
\cpugpu~\cite{aibox,tensordimm} based inference system for personalized
recommendations. As discussed in \sect{sect:sysarch_recsys},  both studies
assume the host CPU memory stores the memory-hungry embedding tables as assumed
in our baseline system architecture.  More relevant to our study is
work by Kwon et al.~\cite{tensordimm} and Ke et al.~\cite{recnmp} which
proposes a near-memory processing architecture for recommendation inference.
As we detail in \sect{sect:time_breakdown}, these prior work is specifically designed as a
point solution to the inference pass of recommendations, so it is not able to
effectively handle the gradient expand-coalesce primitives of embedding layer's
backpropagation step.  To the best of our knowledge, this work is the first to
provide a detailed characterization study on training recommendations,
				revealing important system-level bottlenecks such as gradient
				expand-coalesce or gradient scatter.  More crucially, our treatment of the gradient
				expand-coalesce primitive into a generic tensor gather-reduce
				operator using our \tcasting algorithm is a significant first step
				in developing a training system that comprehensively accelerates the
				forward and backpropagation of embedding layers using a single
				accelerator architecture.

\section{Workload Characterization} 
\label{sect:characterization}

In this paper, we utilize the open-source deep learning recommendation model
(DLRM)~\cite{facebook_dlrm} as a vehicle to conduct a workload characterization
study on training recommendations over \cpuonly and \cpugpu
systems\footnote{Although \emph{training} recommendations 
	commonly employ \cpugpu systems (unlike those for \emph{inference}
			where both \cpuonly and \cpugpu are popular design points), we
		nonetheless study both designs in this section for completeness of
		our characterization.}.  Similar to
		\cite{dlrm:arch,deeprecsys,recnmp}, we explore performance
		acceleration strategies for four model configurations
		(\dlrm{1}/\dlrm{2} as embedding intensive configurations whereas \dlrm{3} and
		 \dlrm{4} are limited by MLPs) representative of two canonical classes of
		recommendation models used for content filtering and ranking.  The chosen
		models attribute to significant ML execution cycles at
		several hyperscaler's datacenters~\cite{deeprecsys,recnmp} and we use them
		to root-cause the performance limiters of training recommendations.  
		Unlike CNNs or RNNs which are trained with a batch size of several
		hundreds, prior work on training DNN-based recommendation models typically
		employ several tens of thousands of mini-batches~\cite{dlrm:github,mlperf:github}. Therefore,
		this paper assumes	a default batch size of $2048$ (the nominal
				batch size of DLRM) but sweeps this number from $1024$ to $4096$ to
		examine sensitivity. \sect{sect:methodology} further details our methodology.

\begin{figure}[t!] \centering
\vspace{-.8em}
\includegraphics[width=0.49\textwidth]{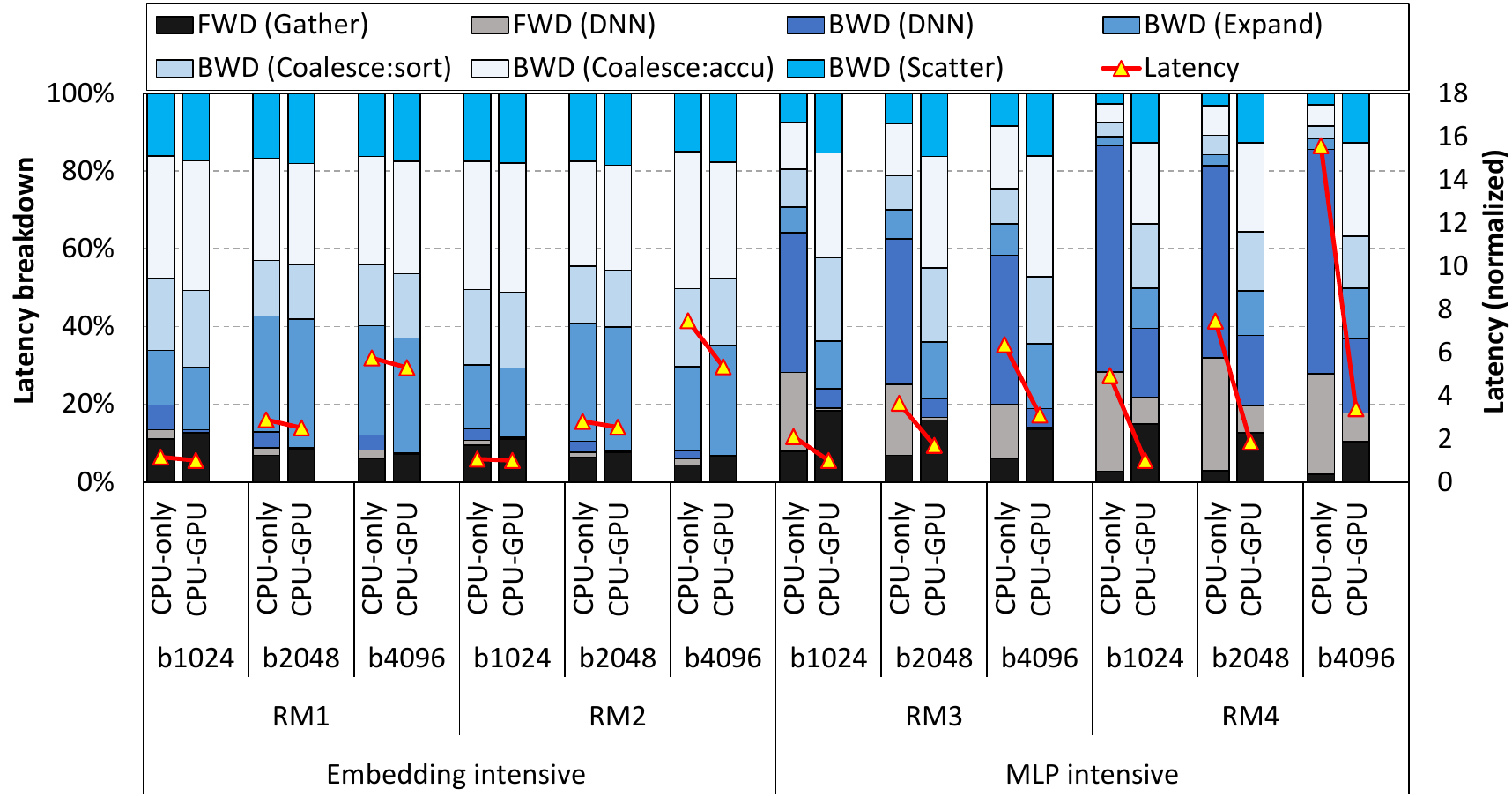}
\caption{
Training time broken down into key steps of forward (black-gray) and backward
	 propagation (blue) as a function of batch size (left-axis). Note that the latency of
	 gradient coalescing (\algo{algo:coalescing}) is broken down into the sorting (\texttt{BWD(Coalesce:sort)})
	and accumulation step (\texttt{BWD(Coalesce:accu)}) for a detailed analysis.
	 The
training time normalized to the fastest configuration of each model (i.e., \cpugpu with
batch $1024$), is shown on the right-axis.
}
\vspace{-1.3em}
\label{fig:latency_breakdown_all}
\end{figure}

\subsection{Breakdown of End-to-End Training Time}
\label{sect:time_breakdown}

\fig{fig:latency_breakdown_all} shows a breakdown of end-to-end training time
(left-axis) as well as normalized latency (right-axis).  Some key observations
that can be made from this figure are as follows.  First, there exists a
noticeable performance gap between \cpuonly and \cpugpu, especially for MLP
intensive \dlrm{3} and \dlrm{4}, highlighting the significant role GPUs play in
training recommendations.  Interestingly however, MLPs	account for only a
small fraction of overall training cycles under \cpugpu (less than 
$1\%$ for embedding limited \dlrm{1}/\dlrm{2} and
		$24\%$ for MLP limited \dlrm{3}/\dlrm{4}), rendering the remaining
embedding layers the most prominent performance bottleneck.  Second, the
backpropagation step of embedding layers accounts for approximately
$62$$-$$92\%$ of end-to-end training time. This shows the
importance of understanding and properly accelerating the backpropagation step
for training recommendations.  Third, the tensor gather-reduce (forward) and
gradient expand-coalesce primitive followed by gradient scatter
(backward) account for the majority of training time, amounting to an average
$60\%$ and $93\%$ for \cpuonly and \cpugpu, respectively. In
particular, the gradient expand-coalesce takes up a  
substantial fraction of backpropagation's latency, causing a significant
bottleneck.

\begin{figure}[t!] \centering
\vspace{-1.7em}
\subfloat[]
{
	\includegraphics[width=0.49\textwidth]{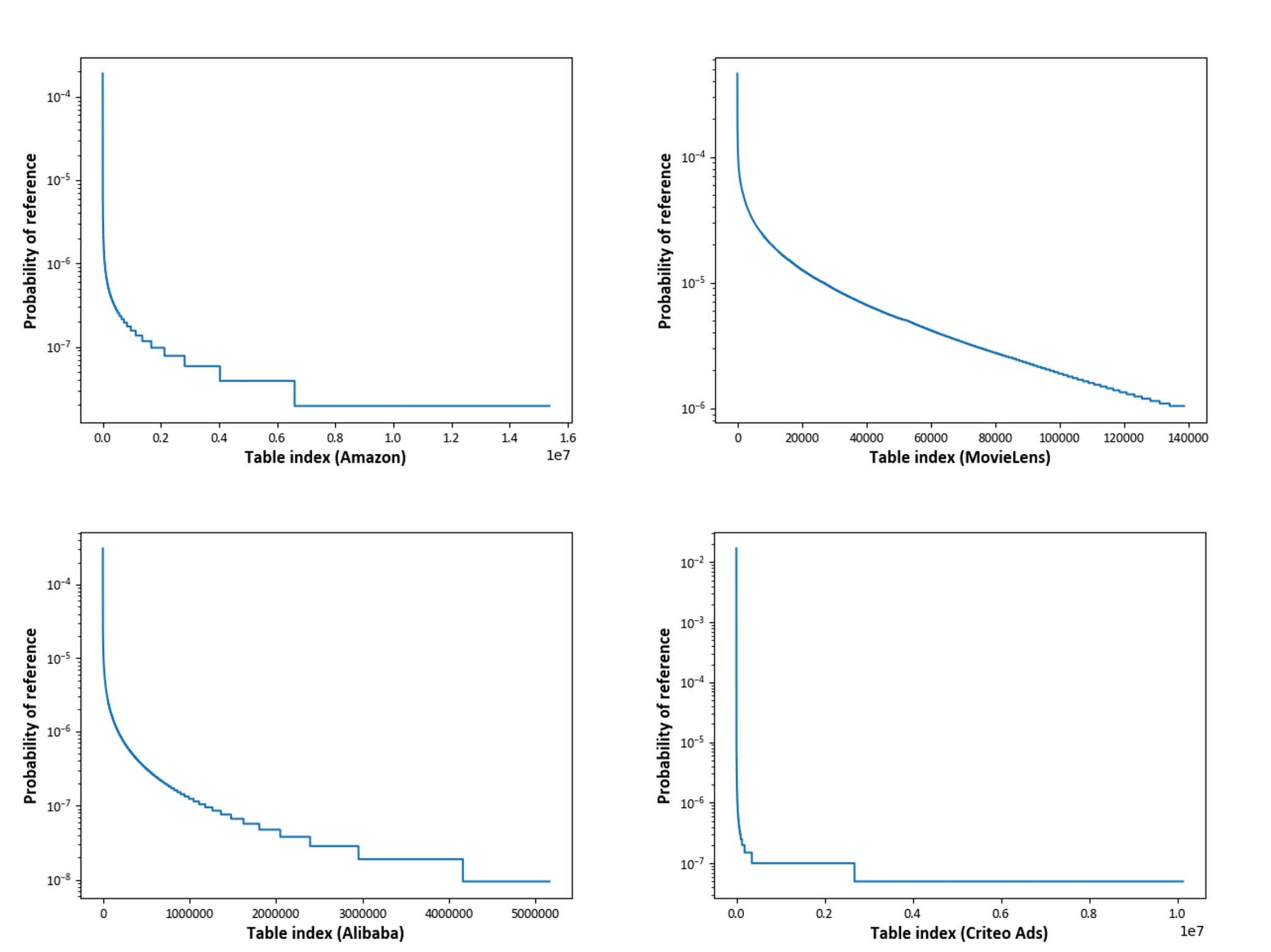}
}
\vspace{-0.8em}
\subfloat[]
{
\includegraphics[width=0.485\textwidth]{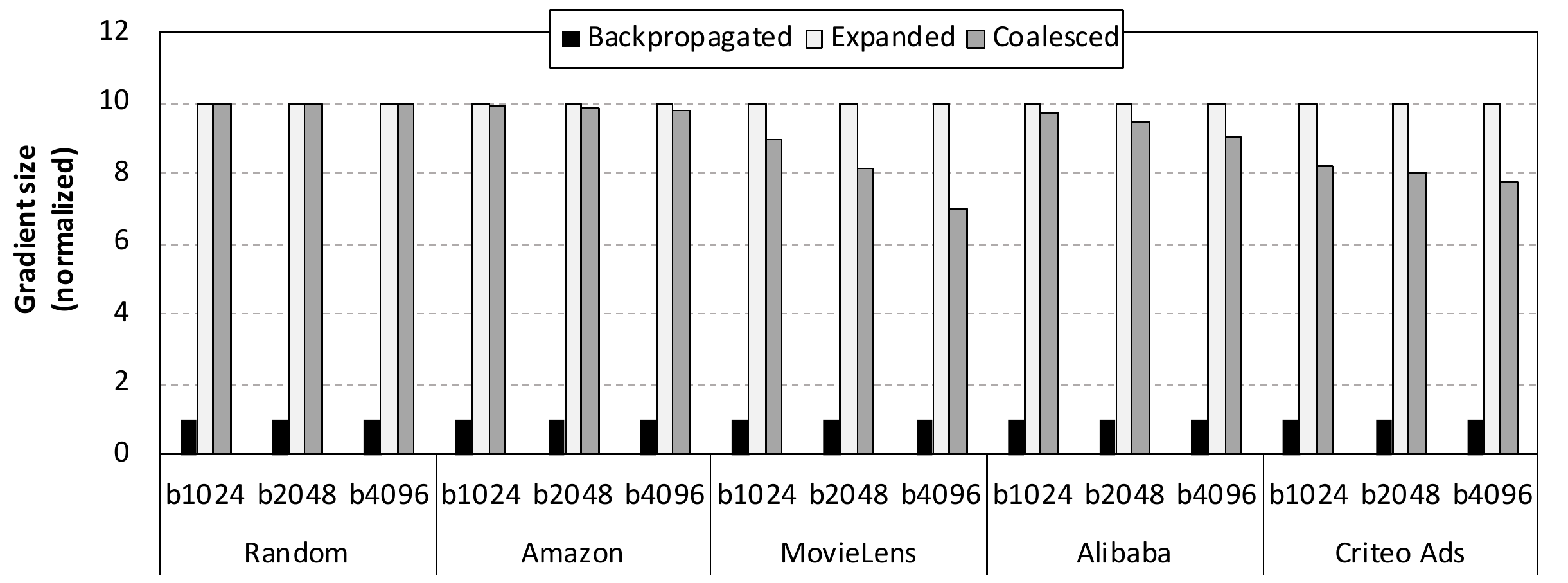}

}
\vspace{-0.5em}
\caption{
	(a) Probability function that quantifies an embedding table entry's likelihood
		of lookup per each dataset. 
		(b) The size of gradient tensors before/after it is expanded and coalesced. 
		Our experiment assumes each table is gathered $10$ times, which is why
		the expanded gradient size is precisely $10\times$ larger than the 
		initial backpropagated gradients. Results are normalized to the size of the backpropagated
		gradient tensor. \emph{Random} assumes a uniform random distribution
		in modeling the probability of embedding table lookups.
}
\vspace{-1.3em}
\label{fig:mem_usage_expand_coalesce}
\end{figure}

\subsection{The Effect of Gradient Expand-Coalesce}
\label{sect:mem_usage_grad_expand_coalesce}

As our characterization uncovered gradient expand-coalesce as causing a
significant performance bottleneck, this subsection provides more detailed
analysis on the important properties of this key primitive in relations to
various recommendation training datasets and the data locality therein.  As
discussed in \fig{fig:recsys_training}, the gradient vectors backpropagated
from the DNN layers must first be expanded and subsequently coalesced before
the coalesced gradients are scattered back to the embedding tables.  Note that
the number of expanded gradients that should be coalesced is a function of how
frequently the \emph{same} sparse index ID (i.e., $src$ ID$=$$2$ in
		\fig{fig:recsys_training}(a)) was used for embedding gathers during forward
propagation.  That is, the more locality observed during the embedding gather
process, the
higher the likelihood of coalescing to reduce the size of the post-coalesced
gradient tensor. To accurately reflect the locality inherent in embedding
gathers for our characterization, we used publicly available training datasets
for recommendations which include Amazon Review (Books)~\cite{amazon_product},
	MovieLens-20M~\cite{movielens}, Alibaba's TaoBao UserBehavior dataset~\cite{alibaba_taobao}, and
		Criteo AI Labs Ad Kaggle~\cite{criteo:dataset} to generate the sparse index IDs
		utilized for embedding table lookups.  Using these publicly available
		datasets, we establish a histogram that counts the number of lookups for
		each distinct index IDs within a given embedding table.  The sorted
		histogram is then utilized to generate the probability function of each
		embedding table entry's likelihood of potential lookups.  The recommendation
		datasets we study contain (several tens of) multiple tables, so for brevity,
		\fig{fig:mem_usage_expand_coalesce}(a) illustrates the probability function
		of the largest embedding table within each dataset.  As depicted, a subset
		of table entries exhibit high access frequencies, which
		highlights the importance of the gradient coalescing step to
		accurately derive the accumulated gradient values for training.  Utilizing
		the probability function in \fig{fig:mem_usage_expand_coalesce}(a), we
		measure the size of the backpropagated gradient vectors before and after
		they are expanded, and subsequently coalesced, the result of which is
		summarized in \fig{fig:mem_usage_expand_coalesce}(b).  As we increase the
		training batch size, the likelihood of sparse indices for gathers
		``hitting'' with each other is gradually increased as more table lookups
		are initiated. Consequently, the effectiveness of expanded gradient's
		getting shrunk through coalescing is gradually increased as batch size gets
		larger. Even after coalescing however, there exists a non-trivial amount
		of coalesced gradient vectors that are subject for scatters, underscoring
		the importance of accurately modeling this crucial step for training.

\begin{figure}[t!] \centering
\vspace{-0.5em}
\includegraphics[width=0.485\textwidth]{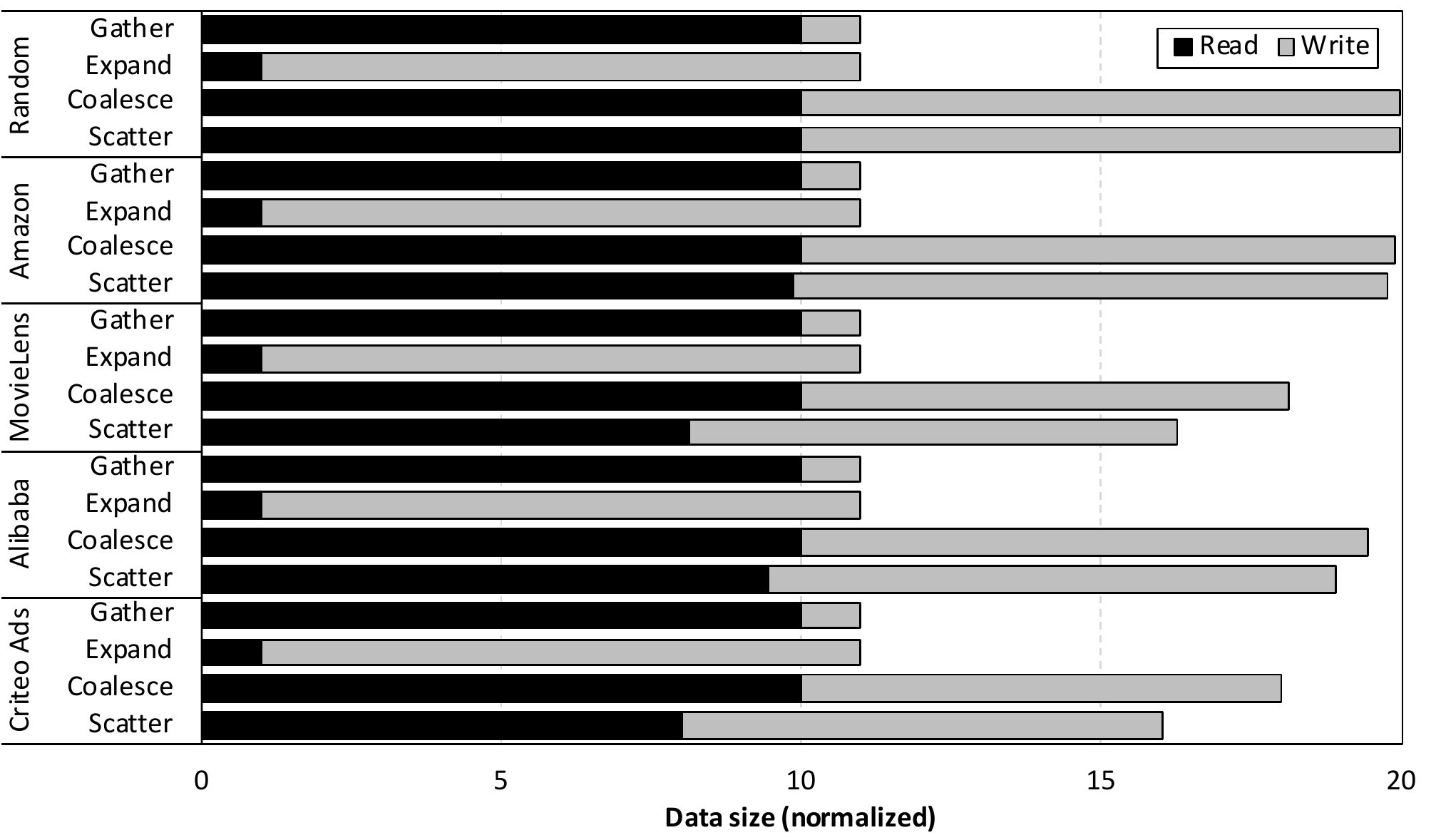}
\vspace{-0.5em}
\caption{
Memory read/write traffic for key primitives in embedding layers. Similar
	to \fig{fig:mem_usage_expand_coalesce}, the analysis assumes each table
	is gathered $10$ times.
}
\vspace{-1.5em}
\label{fig:memory_traffic}
\end{figure}

\subsection{Memory Intensity}  
\label{sect:mem_traffic}

To better understand the high latency overheads of backpropagation,
	 this subsection analyzes the memory throughput demands of
	 embedding layers.  \sect{sect:layer_type} discussed the memory intensive
	 nature of key primitives in training embedding layers. To quantify the
	 \emph{microarchitecture independent} behavior of embedding layer's key
	 primitives, we derive the amount of data the processor loads and stores for
	 each primitive, which can be derived analytically by its algorithmic
	 property.  \fig{fig:memory_traffic} summarizes the memory read/write data
	 traffic demands of the key primitives of embedding layers. Because the
		 memory traffic demands of gradient coalesce is dominated by the gradient accumulation step
			 (and not by the index array sorting step, see \algo{algo:coalescing}), the ``Coalesce'' bar in \fig{fig:memory_traffic} only
	 accounts for the gradient accumulation step within the gradient coalesce operator (i.e., only considers 
			 \texttt{BWD(Coalesce:accu)} and not \texttt{BWD(Coalesce:sort)} in \fig{fig:latency_breakdown_all}).
	 In general,
	 gradient coalesce and gradient scatter incur significantly higher memory
	 traffic than gather-reduce (and correspondingly gradient expand), which explains the large
	 fraction of training time spent conducting backpropagations. In particular,
	 the gradient expand-coalesce step in aggregate incurs an around
	 $3\times$ higher memory traffic than embedding gather-reduce. 
	 Furthermore, the sorting step in gradient coalescing (\algo{algo:coalescing}) adds additional computation overhead
	 on top of the already memory throughput limited gradient accumulation procedure.
	 This explains why gradient expand-coalesce collectively accounts for a
much more significant training time
	 proportion when compared against other
	 memory intensive primitives.  Overall, our workload
	 characterization on training recommendation models root-caused the
	 backpropagation step of embedding layers, specifically the gradient
	 expand-coalesce primitive, as a crucial system-level bottleneck.

\section{Tensor Casting: Co-Designing Algorithm and Architecture for
Recommendation Training}
\label{sect:proposed}

We propose a vertically integrated solution encompassing multiple levels in the computing
stack, ranging over algorithm (\sect{sect:proposed_algo}), software runtime
system (\sect{sect:runtime}) and architecture (\sect{sect:tcast_arch}).
At the heart of our proposal is
our novel \emph{Tensor Casting} algorithm which ``casts'' the gradient expand-coalesce
primitive into a tensor gather-reduce operator. The reason why our casting algorithm
specifically targets tensor gather-reduce is twofold. First,
						 Tensor casted (henceforth referred to as \tcasted) gather-reduce operation algorithmically
						 reduces its memory intensity by $2\times$ compared to
						 gradient expand-coalesce, improving performance.
						 Second, it enables the development of a \emph{generic} accelerator
						 specifically optimized for tensor gather-reduce that encompasses
						 \emph{all} the key primitives of embedding layer
						 training.  In the rest of this paper, we assume \cpugpu
									 as the baseline system as the characterization in
										 \sect{sect:characterization} demonstrated \cpugpu's
										 superior performance compared to \cpuonly.

\subsection{Algorithm}
\label{sect:proposed_algo}

\begin{figure}[t!] \centering
\vspace{-0.5em}
\includegraphics[width=0.485\textwidth]{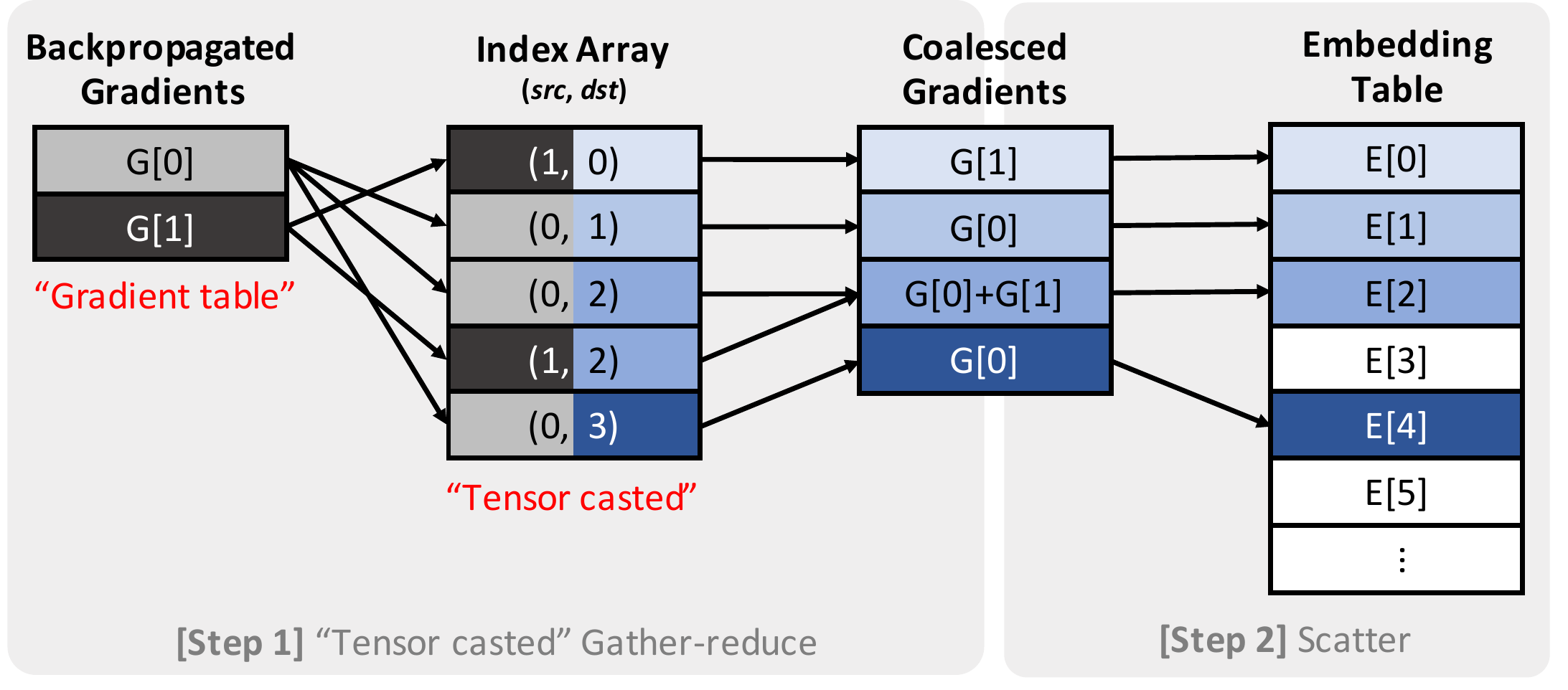}
\vspace{-1em}
\caption{
The baseline gradient expand-coalesce ``Tensor Casted''
into our proposed gradient gather-reduce operation. 
}
\vspace{-1.3em}
\label{fig:tensor_casting_example}
\end{figure}

\begin{figure}[t!] \centering
\vspace{-0.7em}
\includegraphics[width=0.485\textwidth]{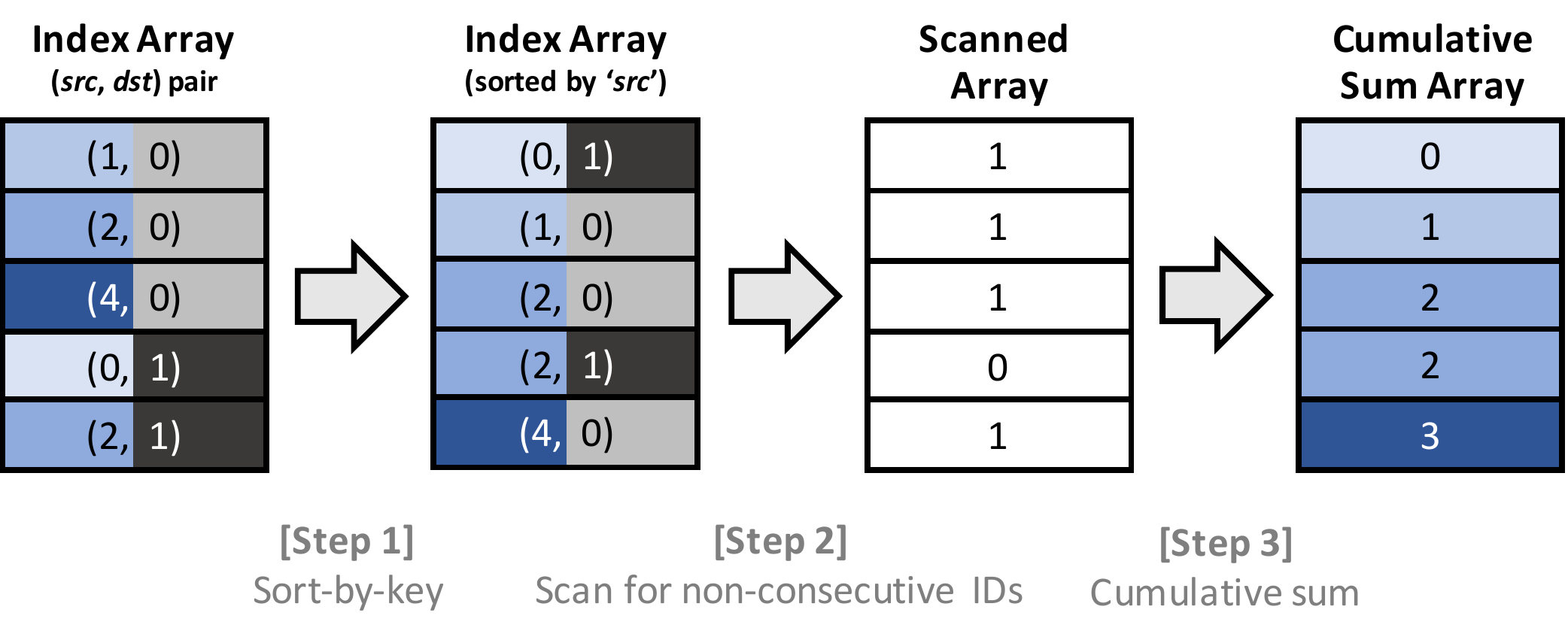}
\vspace{-1em}
\caption{
Illustration of how our \tcasting algorithm derives
the ``\tcasted'' ($src$, $dst$) ID array, which is required by the casted
gradient gather-reduce operation in \fig{fig:tensor_casting_example}. 
}
\vspace{-1.5em}
\label{fig:casting_algorithm}
\end{figure}

{\bf Key observations.} An important observation behind our \tcasting algorithm is
that \emph{coalescing} gradients is functionally equivalent to conducting
\emph{reductions} among the target gradients. If we were to think of the
backpropagated gradient vectors as a ``table'' storing the gradients subject
for coalescing, a gradient expand-coalesce operation is nothing more than
\emph{gathering} the necessary gradient vectors from the ``gradient table'',
	all of which are subsequently \emph{reduced} (i.e., coalesced).
	\fig{fig:tensor_casting_example} provides a high-level overview of the effect of
	\tcasting, which transforms the two-step procedure of gradient
	expand and coalesce (\fig{fig:recsys_training}(b)) into a single, \emph{fused} kernel call
	of \tcasted gather-reduce.  Notice that the \tcasted gradient gather-reduce
	operation requires its own \tcasted ($src$, $dst$) pairs of index array to determine:
	1) which gradient vectors to gather from the ``gradient table'' (using $src$), and
	2) subsequently store the reduced gradients (using $dst$). This \tcasted index
	array is generated during the \emph{casting} stage of our proposed
	algorithm (\sect{sect:runtime} details when/where the casting
			step is conducted), which is detailed in \algo{algo:casting}.  We use
	\fig{fig:casting_algorithm} as a driving example to highlight the important steps
	of \algo{algo:casting}.

{\bf Implementation.}	As discussed in \sect{sect:layer_type}, the gradient
	vectors subject for coalescing are for those embeddings that have been
	gathered-reduced across different batches during forward propagation (e.g., \texttt{E[2]}
			in \fig{fig:recsys_training}). Consequently, the first step in
	\algo{algo:casting} \emph{sorts-by-key} the original ($src$, $dst$) pairs of index
	array using the $src$ ID as the
	$key$ (line $3$ in \algo{algo:casting}). The goal of such sort-by-key
	procedure is to have those $src$ indices gathered more than once during
	forward propagation to appear as consecutive elements within the 
	sorted index array, allowing us to
	determine the \emph{coalescable} indices (i.e., ID=$2$ in the $src$ part
 of the sorted array,	\texttt{[0,1,{\bf 2},{\bf 2},4]}).
	As the $dst$ ID part of the sorted
array pair now designates the batch ID number where 
the gather-reduced embedding vector was stored (see
		\fig{fig:recsys_training}(a)), we can utilize it as the $src$ ID part of the
\tcasted index array (line $4$). In other words, the \tcasted gradient gather-reduce can utilize
the $dst$ ID part of the sorted array as its own \tcasted $src$ ID
to gather the necessary gradient vectors within the 
 ``gradient table'' (i.e., the black-gray colored \texttt{[1,0,0,1,0]} array in
		\fig{fig:tensor_casting_example}).
 As the $src$ part of the sorted index array now stores coalescable indices
 in consecutive index locations, we \emph{scan} the sorted index
array followed by a cumulative sum operation (line $5-9$) to
derive where the gathered-reduced gradients should eventually be stored within
the coalesced gradient vector array. As a result, the final output array in
\fig{fig:casting_algorithm}
can be utilized as
the $dst$ ID part of the \tcasted index array to determine where the reduced
	gradients should be stored (i.e., the \texttt{[0,1,2,2,3]}).

{\bf Merits.} A key advantage of \tcasted gradient gather-reduce is as follows. First, the
two-step procedure of gradient expand and coalesce is now \emph{fused} into a
single gradient gather-reduce, significantly alleviating the high memory
demands of this crucial bottleneck.  Second, our proposal utilizes the
hybrid nature of \cpugpu system to hide the latency
incurred during the casting stage of \tcasting. Note that the baseline
gradient expand-coalesce conducts the sorting step (i.e., the
		process in which the coalescable gradients are determined, \algo{algo:coalescing}) as part of the
gradient coalescing operation. Such implementation renders the latency to determine the
coalescable indices to be directly exposed as part of the backpropagation
latency.  Our unique observation is that the casting stage of \tcasting (\algo{algo:casting}) can
be completely \emph{decoupled} from the \tcasted gradient gather-reduce
operation, allowing us to intelligently hide the casting latency during
forward propagation step. Below, we detail our proposed runtime system that is
co-designed with the underlying system architecture to maximize the benefits of
\tcasting.

\begin{algorithm}[t!]
\small
	\caption{Tensor Casting}
	\label{algo:casting}
	\begin{algorithmic}[1]
	\Procedure{TensorCasting}{$src$, $dst$}
		\State $n$ $\leftarrow$ length($src$)
		\State ($sorted\_src$, $sorted\_dst$) $\leftarrow$ \Call{SortByKey}{$src$, $dst$}
		\State $casted\_src$ $\leftarrow$ $sorted\_dst$
		\For {$i \leftarrow$ 1 to $n$}
			\State $scan$[$i$] $\leftarrow$ ($sorted\_src$[$i$]!=$sorted\_src$[$i$-1])?1:0
		\EndFor
		\State $scan$[0] $\leftarrow$ 1
		\State $casted\_dst$ $\leftarrow$ \Call{CumulativeSum}{$scan$} - 1
		\State \Return ($casted\_src$, $casted\_dst$)
	\EndProcedure
	\end{algorithmic}
	\end{algorithm}

\subsection{Software Runtime System}
\label{sect:runtime}

\begin{figure}[t!] \centering
\vspace{-0.8em}
\subfloat[]
{
	\includegraphics[width=0.485\textwidth]{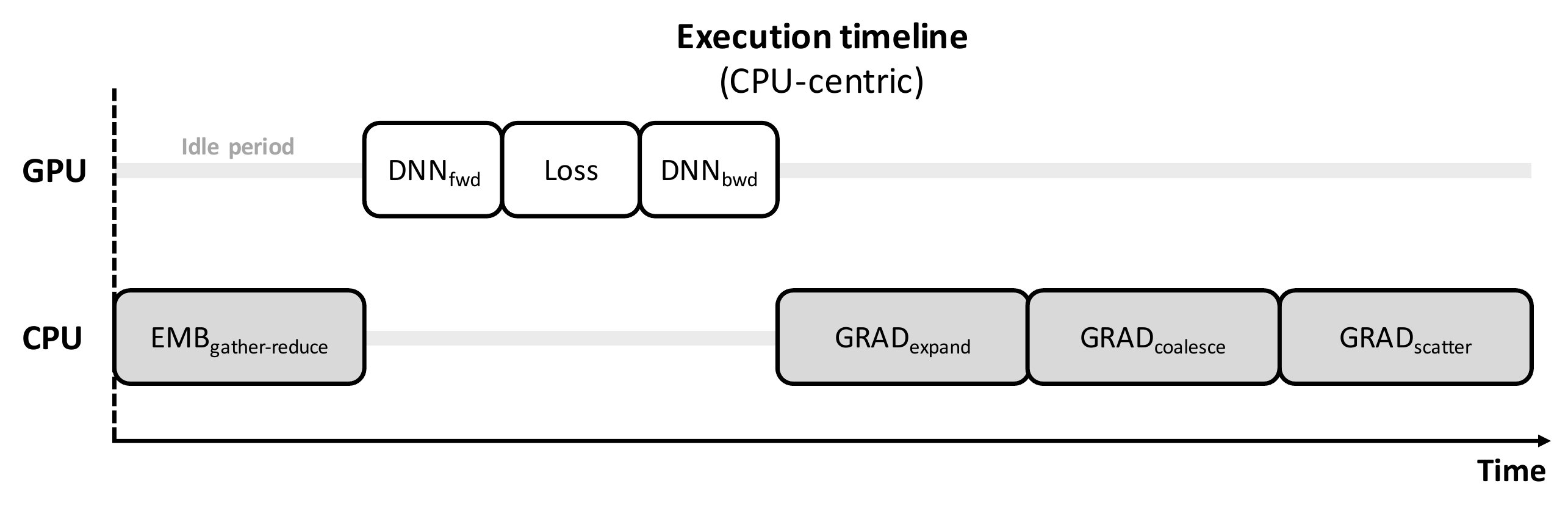}
}
\vspace{-0.5em}
\subfloat[]
{
\includegraphics[width=0.485\textwidth]{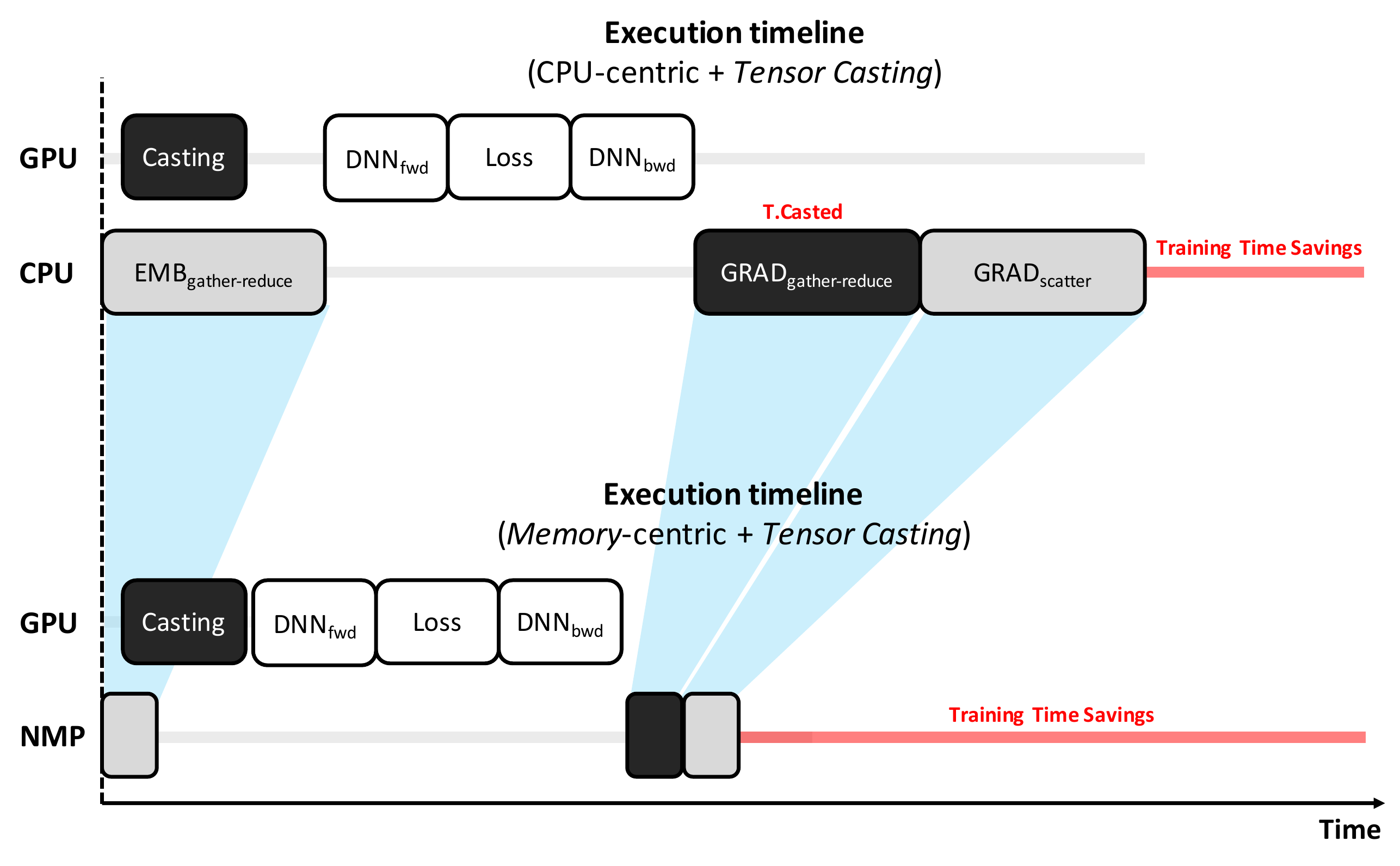}
}
\vspace{-0.1em}
\caption{
Execution timeline of (a) CPU-centric baseline and (b) our proposed CPU-centric
	and memory-centric system using Tensor Casting.
}
\vspace{-1.3em}
\label{fig:execution_timeline}
\end{figure}

{\bf Design principle.} As discussed in \sect{sect:sysarch_recsys}, the conflicting
resource requirements of sparse embedding layers and dense DNN layers mandate a
\emph{hybrid} system architecture for training recommendations: a 
\cpugpu system where the sparse embedding layers are trained on the CPU while the dense DNN
layers are trained on the GPU (\fig{fig:sysarch}). A rather unfortunate
side-effect of such design point is that either one of the processor
architecture remains idle when the other compute engine is busy
executing (\fig{fig:execution_timeline}(a)).  Our software runtime system
utilizes such unique property of hybrid \cpugpu to our benefit by conducting
the \tcasting's casting step (i.e., \algo{algo:casting}) during the forward propagation stage as
depicted in \fig{fig:execution_timeline}(b).  
An important observation from
\algo{algo:casting} is that all the data structures required to generate the
\tcasted ($src$, $dst$) pair of index array is already available at the very beginning of forward
propagation. As the GPU remains idle during the
course of CPU-side embedding gather-reduce, our runtime system copies the original ($src$, $dst$)
	index array to the GPU over PCIe and utilize the GPU to
	proactively initiate the casting step of \tcasting. This allows 
	our runtime system to immediately utilize the \emph{precomputed} \tcasted index array during
	backpropagation and	conduct the \tcasted gradient gather-reduce operation for training time savings.
\algo{algo:castedgatherreduce} summarizes the aforementioned process, which utilizes
the Tensor Casting algorithm (\algo{algo:casting}) to convert the baseline,
		two-step process of gradient
expand-coalescing (\fig{fig:recsys_training}) into a single-step gradient gather-reduce.

{\bf Overhead.} As we detail in \sect{sect:evaluation}, the overhead of copying
the index array is
negligible as its size is only in the order of several MBs, even with several
thousands of input batches.  More importantly, the advantage of such GPU-side
casting operation is that its latency can be hidden inside the CPU-side
embedding gather-reduce operation.  Overall,
					 the proposed runtime system and \tcasting 
					 utilize the currently available software stack as-is, so an
					 important benefit of our proposal is that it can be implemented
					 purely in software over existing CPU-centric \cpugpu systems. As we
					 quantitatively demonstrate in \sect{sect:eval_perf}, \tcasting applied on
					 top of the baseline \cpugpu provides $1.2$$-$$2.8\times$
					 improvement in training throughput.

\subsection{Architecture}
\label{sect:tcast_arch}

We have so far discussed the merits of co-designing the training algorithm
under the CPU-centric \cpugpu systems.  Nonetheless, conventional \cpugpu
systems do not fully reap out the full potential existent with our \tcasting
algorithm, as the memory-limited primitives of embedding layers are still
conducted over the memory-bandwidth limited CPU memory subsystem.  To quantify the
maximum performance improvements possible with our \tcasting, this subsection
presents our \emph{memory-centric} approach for training recommendations.
Before we detail our proposed architecture, let us first discuss some important
properties and requirements of accelerator architectures for training.

{\bf Architectures for training ``dense'' DNNs.}  Because SGD based training
involves the derivation of input and weight gradient vectors, a robust training
architecture should be able to cover the important primitives of both forward
and backward propagation.  Consequently, TPUs for training
are typically optimized for GEMM operators because of the versatility of this
all-round compute primitive~\cite{volta_v100,tpu2,gaudi,ipu,cerebras}.
Concretely, the compute primitives of both forward and backward propagation of
popular dense DNN layers (e.g., convolutional, fully-connected, and recurrent
		layers) are able to be ``casted'' as GEMMs, amortizing the design cost of
TPUs by specializing its microarchitecture for GEMMs.  As such, an
important design principle in TPU architectures for training DNNs is
maintaining its applicability to key primitives of both
forward and backward propagation while also achieving the high
energy-efficiency of specialized accelerators.

\begin{algorithm}[t!]
	\small
	\caption{Tensor Casted Gradient Gather-Reduce}
	\label{algo:castedgatherreduce}
	\begin{algorithmic}[1]
	\Procedure{GatherReduce}{$src$, $dst$, $grad$}
		\State $n$ $\leftarrow$ length($src$)
		\For {$i$ $\leftarrow$ 0 to $n$}
			\State $coal\_grad$[$dst$[$i$]] $\leftarrow$ $coal\_grad$[$dst$[$i$]] + $grad$[$src$[$i$]]
		\EndFor
		\State \Return $coal\_grad$
	\EndProcedure
		\\
	\Procedure{T.CastedGradGatherReduce}{$src$, $dst$, $grad$}
		\LineComment{Step A: Execute Tensor Casting alogrithm (Algorithm~\ref{algo:casting})}
		\State ($casted\_src$, $casted\_dst$) $\leftarrow$ \Call{TensorCasting}{$src$, $dst$}
		\LineComment{Step B: Initiate GatherReduce kernel for Gradient Updates}
		\State \Return {GatherReduce}{($casted\_src$, $casted\_dst$, $grad$)}
	\EndProcedure
	\end{algorithmic}
\end{algorithm}

\begin{figure}[t!] \centering
	\includegraphics[width=0.39\textwidth]{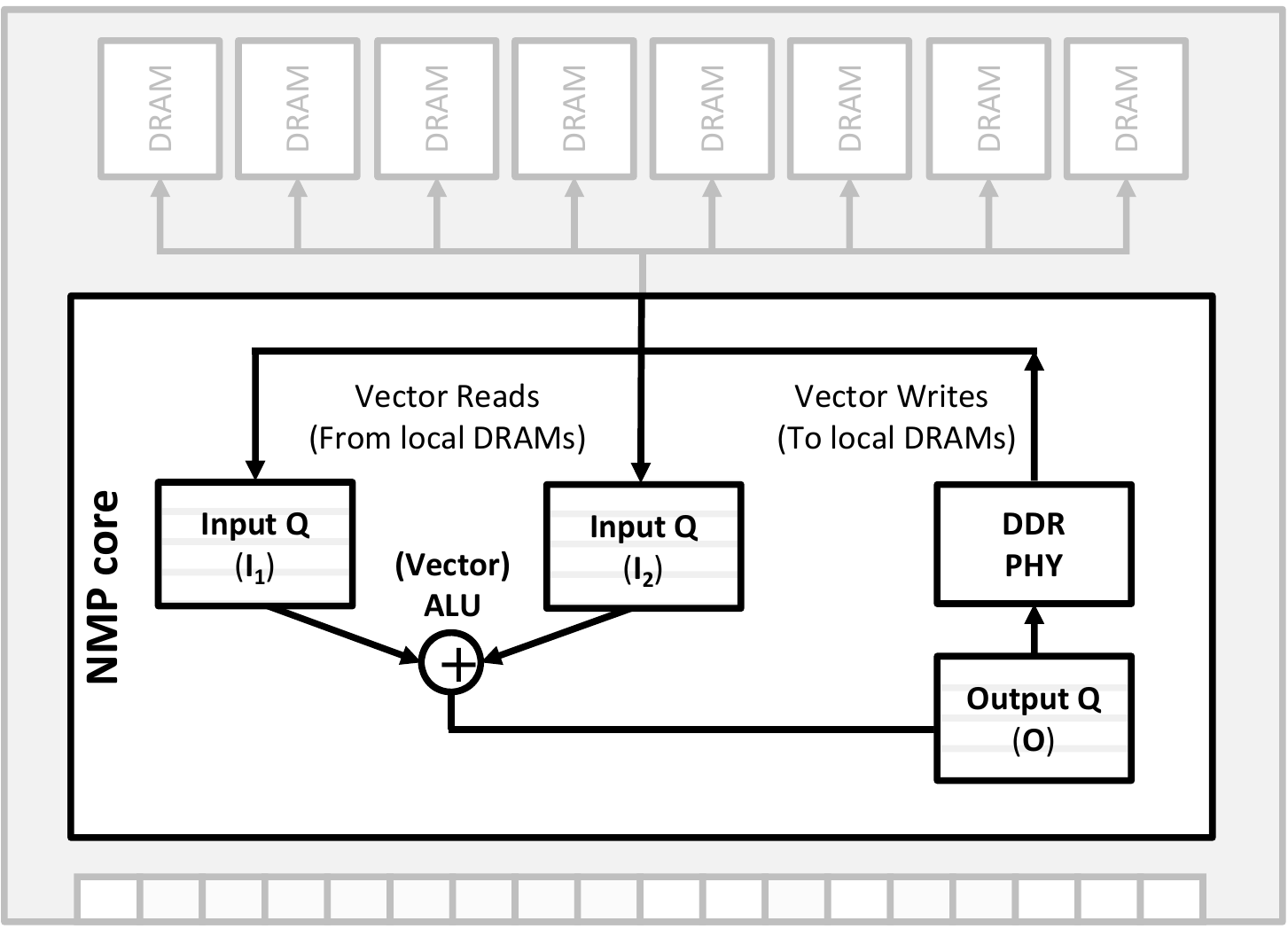}
\caption{
NMP microarchitecture for tensor gather-scatter.
}
\vspace{-1.5em}
\label{fig:nmp_uarch}
\end{figure}

{\bf Architectures for training ``sparse'' embeddings.} Our workload
characterization has root-caused embedding gather-reduce (forward), gradient
expand-coalesce, and gradient scatter (backward propagation) as the
three most time-consuming primitives of training embedding layers.  Given the increasing
importance of this memory-limited ML algorithm and the significant performance
bottleneck it incurs, we argue that ML workloads utilizing embeddings deserve a
targeted acceleration treatment.  \tcasting has been carefully designed from
the ground up to enable the design of a \emph{generic} accelerator architecture
meeting the aforementioned two design requirements of training: its 
applicability to both forward and backpropagation and delivering high-energy efficiency
through specialized microarchitecture designs.  Recall that \tcasting can
transform the gradient expand-coalesce primitives into gradient
gather-reductions. Additionally, notice that the datapath of scatter operations
is virtually identical to gathers as both operations can be conducted over the
same datapath, just in the opposite directions (detailed later using
\fig{fig:nmp_uarch}). As such, the development of our \tcasting algorithm to
transform gradient expand-coalesce into gather-reduce is no coincidence, 
					as we seek to \emph{unify} all key primitives
of training embeddings under a \emph{single} compute operator.
Consequently, our proposition is to optimize the
accelerator microarchitecture using \emph{near-memory processing} (NMP) technology 
for high-throughput tensor gather-reductions as well as tensor scatters.
Building on
top of recent DIMM-based NMP accelerator
designs~\cite{chameleon,alian2018application, alian2019netdimm, tensordimm,recnmp}, the {\bf key
	innovation} of our proposal is the utilization of the \emph{expressive power
		of the tensor gather-reductions} to architect a sparse-optimized
		accelerator for training embedding layers.  The merits of an NMP-based
		tensor gather-scatter accelerator are clear: 1) the design can cover the
		majority of embedding training time using a single microarchitecture, and
		2) the NMP design paradigm can fundamentally address the memory throughput
		bottlenecks of embedding layers. 

\begin{figure}[t!] \centering
	\includegraphics[width=0.43\textwidth]{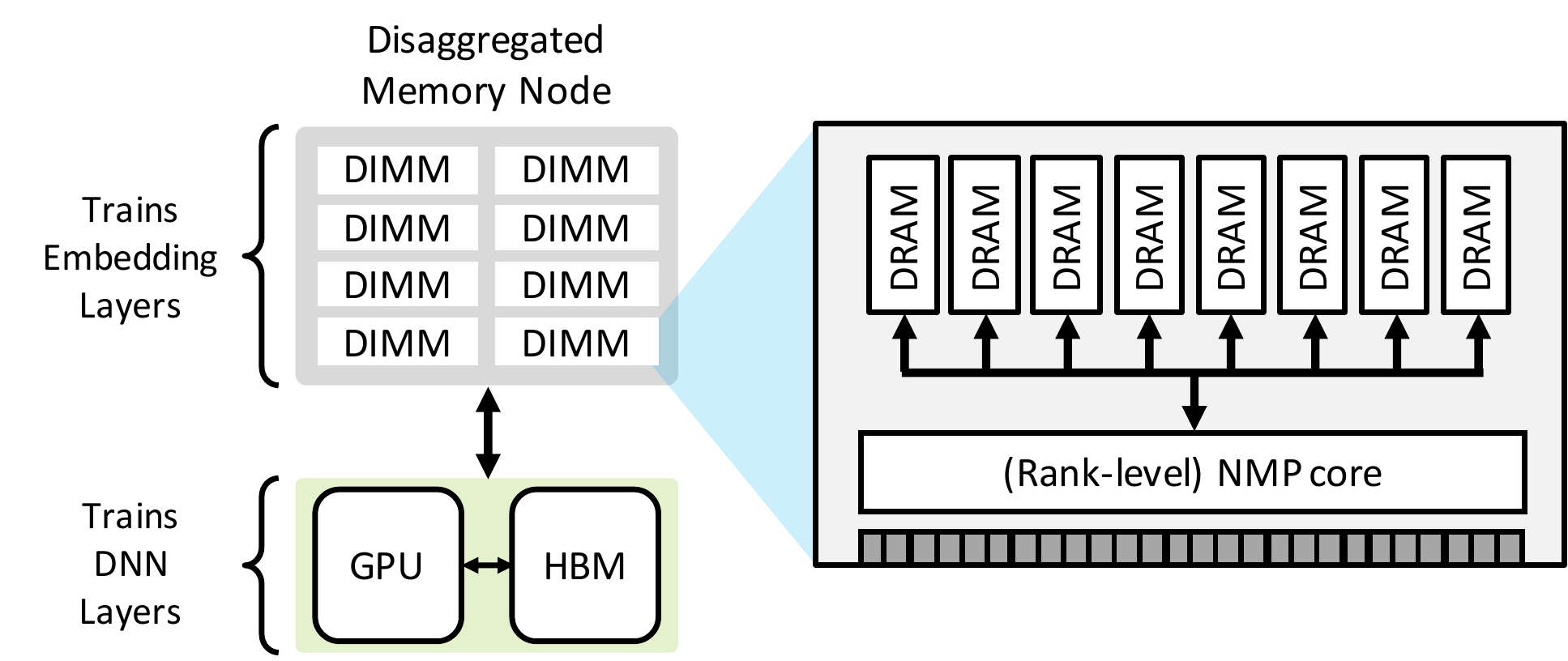}
\caption{
Proposed memory-centric system for training recommendations.
}
\vspace{-1.5em}
\label{fig:nmp_sysarch}
\end{figure}

{\bf A ``memory-centric'' recommendation training.} There have been several
prior studies~\cite{chameleon,alian2018application,alian2019netdimm,tensordimm,recnmp} 
exploring the advantages of augmenting commodity DIMM devices with NMP
accelerators optimized for specific application domains. In particular,
prior work by Kwon et
al.~\cite{tensordimm} and Ke et al.~\cite{recnmp} each explored the merits of
NMP acceleration for \cpugpu and \cpuonly based systems for recommendation
inference.  As discussed in \sect{sect:sysarch_recsys}, systems for training
typically utilize GPU for accelerating dense DNN layers, so we discuss how \tcasting can be
employed on top of the GPU-centric disaggregated memory system as suggested by
Kwon et al.  Nonetheless, \tcasting can be readily employed under the
CPU-centric NMP accelerator proposed by Ke et al. because the key intuition behind
our proposal (i.e., utilizing the expressiveness of tensor gather-reduce for
		accelerating both forward/backward propagation) is orthogonal to the
underlying CPU vs. GPU based NMP microarchitecture.

		\fig{fig:nmp_sysarch} provides a high-level overview of our memory-centric
		approach in training recommendations.  Here, the acceleration of sparse
		embeddings and the dense DNNs is separated across the
		disaggregated, pooled memory architecture and the GPU,
		respectively.  The disaggregated memory pool is populated with multiple
		units of custom designed DIMMs, each of which is augmented with a NMP
		accelerator to be able to \emph{locally} train the embedding tables. The aggregate
		memory capacity provided with the pool memory architecture can be several
		TBs when utilizing the latest capacity-optimized
		LR-DIMMs~\cite{fb:zion,lrdimm}, allowing the entire embedding tables to be stored
		locally.  For training DNNs, the input tensor is copied over to the
		GPU to leverage its abundant compute and memory throughput.
		In terms of the NMP accelerator microarchitecture, our proposed design builds 
		on top of the work by Kwon et al.~\cite{tensordimm} and Ke et al.~\cite{recnmp}
		as detailed below.

{\bf NMP microarchitecture.} \fig{fig:nmp_uarch} details our NMP core
design, each with 1) a vector ALU conducting the
reduction among the gathered embeddings, 2) a local memory controller
that translates the tensor gather-reduce and scatter instructions into
low-level DRAM commands, and 3) a set of input/output buffers
that stage in/out the embedding vectors or other metadata required for tensor
gather-scatter. Similar to the design suggested by
Kwon et al.~\cite{tensordimm}, we assume that the GPU sends a CISC instruction
encapsulating the necessary information required to conduct tensor gather-reduce (and
similarly scatter), which the NMP core receives to conduct the necessary
transactions locally within the DIMM.
To support fine-grained embedding
gather and scatter operations, the NMP core is equipped per each rank to utilize
rank-level parallelism for bandwidth amplification. As the minimum access granularity
per each rank is $64$ bytes, each NMP core is able to conduct multiples of
$64$ byte granularity gathers and scatters.
The multiple embedding tables are
interleaved across the ranks such that the \emph{effective} memory throughput
available across the NMP cores are amplified as a function of the number of
ranks employed within the disaggregated memory.
 The NMP local
input/output buffers are utilized to conduct an on-the-fly reductions
or temporarily store the data read out of DRAM, the result of which can be drained
back to memory to store the reduced embeddings or for scatter operations. 
Because the NMP-based tensor gather-scatter is conducted without having to move the data
outside the DIMMs, our solution enables significant increase in the aggregate effective bandwidth than
conventional pin-limited memory subsystems. Further details regarding the design
of an NMP unit for gather-reduce, the required system software support and API extensions
are available at \cite{tensordimm,recnmp} and we refer the interested readers to these prior work due to space constraints.

\subsection{Putting Everything Together}
\label{sect:tcast_uarch}

This section described the key facets of our algorithm and
architecture co-design which targets recommendation training systems. Key
innovation of our proposal is our novel \tcasting algorithm, which enables the
acceleration of training embedding tables using a \emph{single} compute
primitive -- tensor gather-reductions.  Based on such insight, we propose,
					implement, and demonstrate our proposal on top of an existing
					CPU-centric system as well as a more future-looking memory-centric
					system architecture, achieving significant training throughput
					improvements (\fig{fig:execution_timeline}).

\section{Methodology}
\label{sect:methodology}

\begin{table}[t!]
\centering
\caption{Disaggregated memory architecture configuration.}
\scriptsize
\vspace{-0.5em}
  \begin{tabular}{|c|c|}
		\hline
    DRAM specification     			&  DDR4 \\
		\hline
    Number of ranks	     			&   $32$ \\
		\hline
    Effective memory bandwidth (per rank)     			& $25.6$ GB/sec   \\
		\hline
    Effective Memory bandwidth (in aggregate)  			& $819.2$ GB/sec   \\
    \hline
  \end{tabular}
\vspace{-1.3em}
  \label{tab:dram_config}
\end{table}

{\bf Evaluation framework.} Our software runtime is designed using the
open-sourced PyTorch backend library (v$1.5.0$) for modeling embedding layers and NVIDIA's
cuDNN/cuBLAS/Thrust~\cite{cudnn,cublas,thrust} and Intel MKL~\cite{mkl} for DNN layers. To
establish a strong baseline to compare against our proposal, we characterize
the performance of the key primitives of PyTorch library and heavily optimize
and tune its performance whenever necessary, for a conservative analysis. For instance, our tuned version of
		gradient coalesce achieves $5.0-6.1\times$ 
		and $6.7-12\times$ higher
		performance than baseline PyTorch for the sorting
		and accumulation steps of gradient coalescing (\algo{algo:coalescing}), respectively,
		by better parallelizing and tuning its execution\footnote{TensorFlow's implementation of the gradient expand-coalesce operator
		exhibited similar performance bottlenecks, so we utilize our optimized/tuned version of PyTorch as baseline for a conservative analysis.}. All the evaluation conducted in
		this paper (including our motivational data in \fig{fig:latency_breakdown_all}) utilizes our optimized implementation of PyTorch. As
\tcasting is purely an algorithmic innovation which can be implemented
completely in software, we utilize CUDA as well as existing software libraries
(e.g., NVDIA's NVlabs CUB~\cite{cub}) and PyTorch runtime APIs
to design our proposed algorithm. We thoroughly validate the functional equivalence between the baseline gradient
expand-coalesce primitive and our proposed tensor casted gradient gather-reduce operator.

As our CPU-centric system with \tcasting can be evaluated over real systems, we
measure the end-to-end wall clock time for reporting performance. To quantify
the benefits provided with our NMP based memory-centric system, we follow the
methodology suggested by Kwon et al.~\cite{tensordimm}  which employs an emulation based study. As the key primitives
of embedding layers (e.g., tensor gather-reduce) are memory bandwidth limited,
	 these prior work utilize a cycle-level DRAM simulator to
	 measure the effective memory throughput of the memory system when
	 fed in with the appropriate DRAM commands. 	 
	 The effective memory
	 throughput is then utilized as a proxy to emulate the performance of the
	 configured memory subsystem when executing the NMP operations over real
	 systems as follows. 	 Our emulated disaggregated memory architecture is 
	 assumed as employing enough number of ranks to deliver an aggregate
	 effective memory bandwidth commensurate to that of a GPU's local HBM memory
	 bandwidth (\tab{tab:dram_config}). 
	 We then use Ramulator~\cite{ramulator} to model
	 our emulated disaggregated memory architecture, which achieves significant
	 memory throughput thanks to its near-memory processing nature (over $600$ GB/sec 
			 of effective throughput over the maximum $819.2$ GB/sec).
	 The NMP-enhanced tensor gather-reduce
	 and scatter functions (\fig{fig:execution_timeline}(b)) are then implemented
	 as a CUDA kernel that emulates the behavior of NMP operations over a real
	 GPU (one which is assumed as the disaggregated memory pool), which we
	 integrate into our software runtime system to measure end-to-end performance
	 of training recommendations. In general, our emulation methodology follows
	 that of \cite{tensordimm,recnmp}.

\begin{table}[t!]
  \centering
  \caption{Recommendation model configurations.}
\scriptsize
\vspace{-0.5em}
  \begin{tabular}{|c|c|c|c|c|}
    \hline
    \textbf{Model} & \textbf{\# of Tables} & \textbf{Gathers/table}& \textbf{Bottom MLP} & \textbf{Top MLP} \\
    \hline
    \hline
    \dlrm{1}						&		10		& 80	 & 256-128-64		& 256-64-1\\
    \hline	                                                       
		\dlrm{2}						&		40	& 80	 & 256-128-64	& 512-128-1\\
    \hline                                                         
    \dlrm{3}   					&   10		& 20	 & 2560-512-64		& 512-128-1\\
    \hline                                                         
    \dlrm{4}   					&   10	& 20		& 2560-1024-64	& 2048-2048-1024-1\\
    \hline
  \end{tabular}
\vspace{-1.3em}
  \label{tab:benchmarks}
\end{table}

{\bf System configuration.} We utilize NVIDIA's DGX system~\cite{dgx_1v}
equipped with eight V100 GPUs~\cite{volta_v100} for our study.  The NVIDIA V100
comes with $900$ GB/sec of memory throughput (similar to our $32$ ranked
		disaggregated memory, \tab{tab:dram_config}) with six NVLINKs for
communicating with the other GPUs up to $150$ GB/sec. A single GPU card
communicating with the host CPU over PCIe(gen3) is assumed when evaluating
the \cpugpu baseline and our \tcasting \cpugpu. For our memory-centric system, we
utilize a pair of V100s to model the \nmpgpu system, where one of the GPUs
emulates the behavior of our NMP-augmented disaggregated memory node.  We
configure the communication bandwidth between \nmpgpu to be $25$ GB/sec, which
is closest we could configure to be commensurate to the PCIe(gen3) bus
bandwidth utilized for \cpugpu. As we detail in \sect{sect:sensitivity}, the
performance of \nmpgpu is insensitive to the communication bandwidth.

\begin{figure*}[t!] \centering
\includegraphics[width=0.985\textwidth]{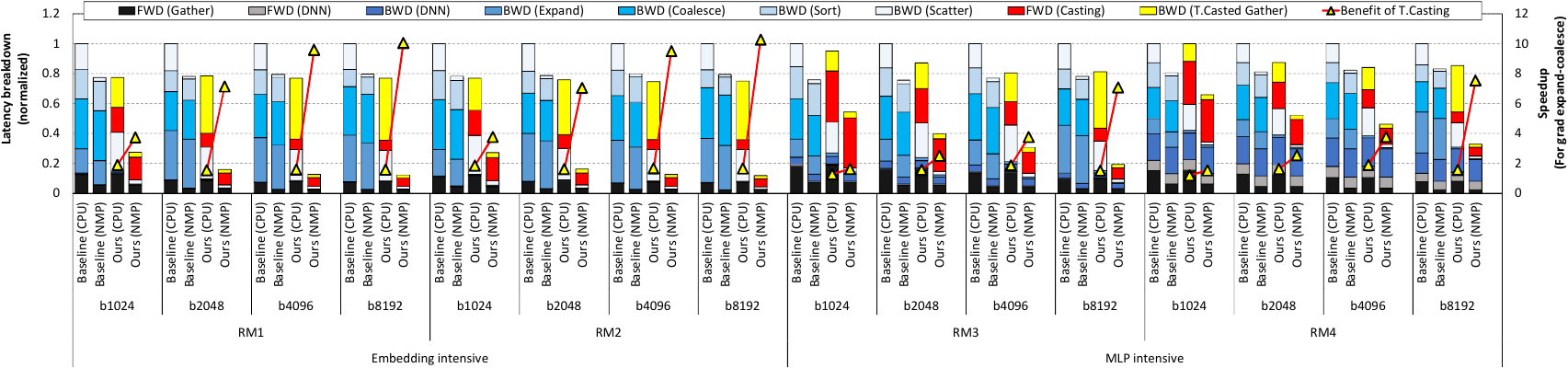}
\caption{
Latency breakdown of our studied workloads. The left-axis shows a stacked bar
	chart of the execution time of each individual compute primitives (normalized to \baseline).
	The right-axis shows the speedup \tcasting brings about just for the gradient expand-coalesce
	step by measuring (execution time of gradient expand-coalesce)/(execution time of the casting step (red) and \tcasted gather-reduce (yellow)).
\texttt{Baseline(NMP)} models a TensorDIMM based NMP design which can accelerate ``both'' embedding gather-reduce and gradient scatters, but gradient expand-coalesce is executed without Tensor Casting, identically as done in \texttt{Baseline(CPU)}.
}
\vspace{-0.7em}
\label{fig:eval_latency}
\end{figure*}


{\bf Benchmarks.} We study four recommendation models using the open-sourced
DLRM~\cite{facebook_dlrm} (\tab{tab:benchmarks}).  \dlrm{1-3} are configured
identically as discussed by Gupta et al.~\cite{deeprecsys}, with \dlrm{1-2}
exhibiting embedding intensive and \dlrm{3} showing MLP intensive behavior.
\dlrm{4} is modeled as an MLP intensive workload by stacking one additional MLP
layer and increasing all MLP layer's parameters.  Following prior
work~\cite{facebook_dlrm,mlperf:github}, the default embedding
vector size is set as a $64$-dimensional vector.  As recommendations are
typically trained with a batch size of several thousands to tens of thousands
of inputs, we study a batch size of $1024$/$2048$/$4096$ as our default
setting.  In \sect{sect:sensitivity}, we study the sensitivity of \tcasting
when deviating from these default configurations.

\section {Evaluation} \label{sect:evaluation}

	We explore four system design points: the 1) baseline CPU-centric system
	utilizing CPU-GPU	as-is (\baseline), 2) TensorDIMM-based baseline NMP accelerator
	which can accelerate both embedding
	gather-reduce and gradient scatters, but gradient expand-coalesce being
	executed identically as baseline CPU-centric system without Tensor Casting (\texttt{Baseline(NMP)}), 
	3) our \tcasting applied CPU-centric system (\ourscc), and
	4) our proposed memory-centric system utilizing both NMP and Tensor Casting (\oursmc). 
	Note that \tcasting \emph{does not change
		the algorithmic nature of SGD training} as our proposal does not
		change the mathematical property of gradient coalescing. Hence, the total
		number of training iterations required to reach the same level of
		CTR accuracy of a recommendation is \emph{identical} between baseline and our
		proposal.

\subsection{Latency Breakdown}
\label{sect:eval_latency}

\begin{figure*}[t!] \centering
\includegraphics[width=0.985\textwidth]{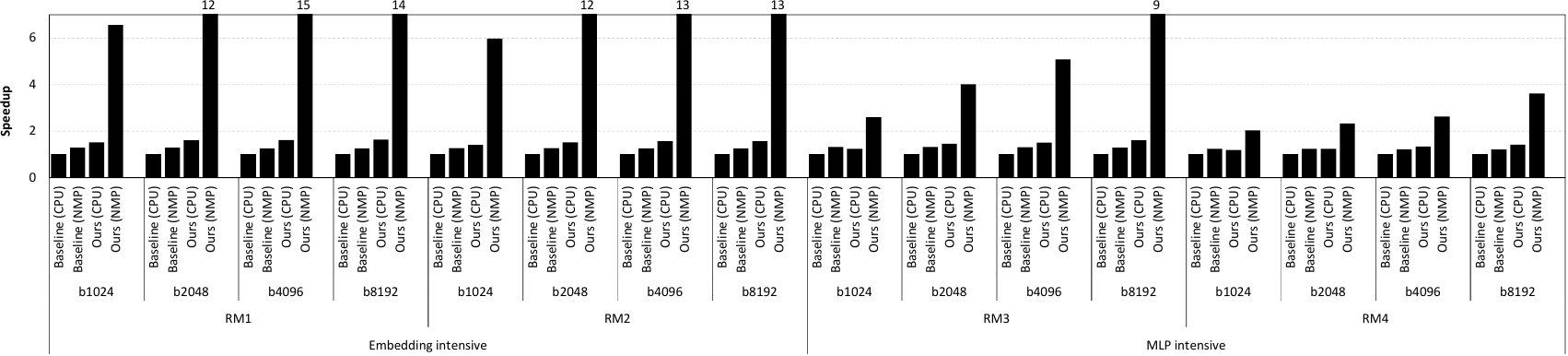}
\caption{
End-to-end performance speedup provided with Tensor Casting.
}
\vspace{-1em}
\label{fig:eval_perf}
\end{figure*}

\begin{figure*}[t!] \centering
	\includegraphics[width=0.985\textwidth]{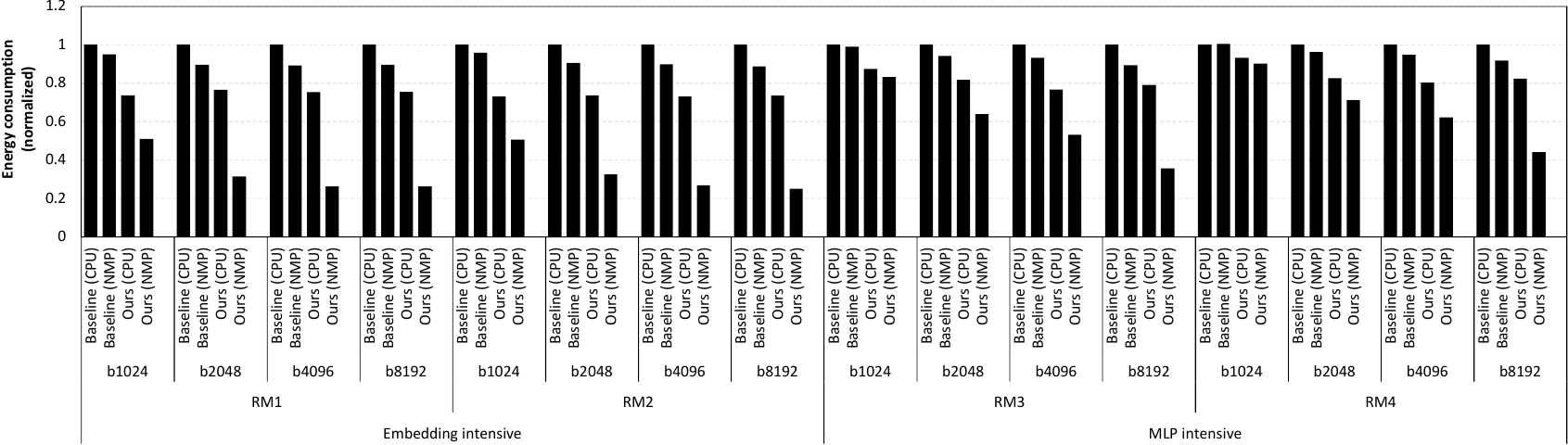}
	\caption{
			Energy consumption with/without Tensor Casting. 
	}
	\vspace{-0.7em}
	\label{fig:energy_eval}
\end{figure*}

As discussed in \sect{sect:runtime}, our software runtime intelligently hides
the casting overhead of \tcasting by initiating this process during
forward propagation. Before we discuss the end-to-end performance speedup
provided with \tcasting, this subsection first quantifies the efficacy of
\tcasting in addressing the performance bottlenecks of the baseline gradient
expand-coalesce primitive.  \fig{fig:eval_latency} shows a latency breakdown of
our studied workloads, which is a stacked bar chart of the execution time of
each individual compute primitives during training (left-axis).  While the
accumulated latency is indicative of performance, it does not exactly reflect
the end-to-end training time as our runtime can hide the latency of the
casting stage of \tcasting. As discussed in \sect{sect:time_breakdown},
							the baseline \cpugpu consumes significant latency during the
							backpropagation stage, with the gradient expand-coalesce
							operator incurring the most noticeable latency overhead.
							\tcasting substantially reduces the time taken to conduct this
							time-consuming step by permuting it into gather-reduce, which is
							represented by the much smaller time taken to execute the
							casting (red) and the actual \tcasted gather-reduce
							(yellow), achieving $1.1-9.5\times$ latency reduction for this
							bottleneck operator (right-axis).  As we detail in
							\sect{sect:eval_perf}, the end-to-end training time reduction is
							even higher as our runtime system hides the latency of
							the casting stage during forward propagation.  As for our 
							memory-centric system, the time taken to conduct the \tcasted
							gradient gather-reduce operation is further reduced by an
							additional $1.3-6.1\times$ 
							compared to \ourscc ($1.5-9.5\times$ speedup than \baseline)
	all thanks to our versatile NMP cores.  
	In fact, the performance advantage
	of NMP is so pronounced that the casting stage can sometimes become a new
	performance bottleneck under our memory-centric system (i.e., the latency
		overhead of casting is identical between \ourscc and \oursmc).

\subsection{System-level Performance}
\label{sect:eval_perf}

\fig{fig:eval_perf} summarizes the end-to-end speedup 
offered with our proposal. In general, as \tcasting primarily
improves the performance bottlenecks of embedding layers, the achieved speedups are
more pronounced on embedding intensive \dlrm{1-2} 
than the MLP intensive \dlrm{3-4}. 
Our software-only \tcasting running on top of hybrid
\cpugpu provide $1.2-1.6\times$ speedup under our default configurations,
with $1.4-2.8\times$ speedup when trained with larger mini-batch sizes (\sect{sect:sensitivity}).
It is worth pointing out that our software-only Tensor Casting (\ourscc) performs even
better than the baseline TensorDIMM-based NMP accelerator (\texttt{Baseline(NMP)}), achieving an average
15\% speedup. While the baseline NMP can utilize near-memory acceleration
to reduce latency in conducting embedding gather-reduce and gradient scatter,
	 the bottleneck incurred by gradient expand-coalesce still remains
	 as the biggest performance limiter. 
These results
	highlight the importance of properly addressing the critical system-level
	challenges of embedding layer's backpropagation step, which our Tensor Casting 
	algorithm successfully delivers.

Under our memory-centric system design, all the key primitives of embedding
layer training (i.e., embedding gather-reduce, gradient expand-coalesce, and gradient
		scatter) achieve significant latency reduction as our co-designed algorithm (\tcasting)
and architecture (NMP cores) successfully accelerates these operations under
our ``unified'' tensor gather-scatter unit. Overall, our memory-centric \tcasting 
approach achieves   
$2.0-15\times$ (average $6.9\times$) training throughput increase,
significantly improving the state-of-the-art in training recommendation models.

\subsection{Design Overheads, Energy-Efficiency, and NMP Utilization}
\label{sect:impl_overhead}

{\bf Design overheads.}
Our software-only
\tcasting algorithm is implemented purely on top of the existing
hardware/software and can readily be deployed over real silicon. As our
CPU-centric \ourscc 
operates
over real systems, the training time reduction directly translate 
into energy-efficiency (shown in \fig{fig:energy_eval}).
Our NMP microarchitecture builds upon
the DIMM-based NMP substrate as proposed in~\cite{tensordimm,recnmp}.
Compared to these prior works, the modifications
required in the DIMM or the underlying NMP microarchitecture
is practically negligible as the key innovation of \tcasting is our permutation
algorithm itself that enable \emph{all} major compute primitives of training
embeddings to operate over the tensor gather-scatter accelerator. 
The primary change required is the inclusion 
of the tensor scatter instruction as part of the ISA.
We implement and synthesized
the major components of our NMP core 
on top of a Xilinx Virtex FPGA board using
Verilog HDL, confirming that the added area and power overheads of the
NMP unit is practically negligible, which is in line with the conclusions
from prior works~\cite{tensordimm,recnmp}.

\begin{figure}[t!] \centering
\includegraphics[width=0.495\textwidth]{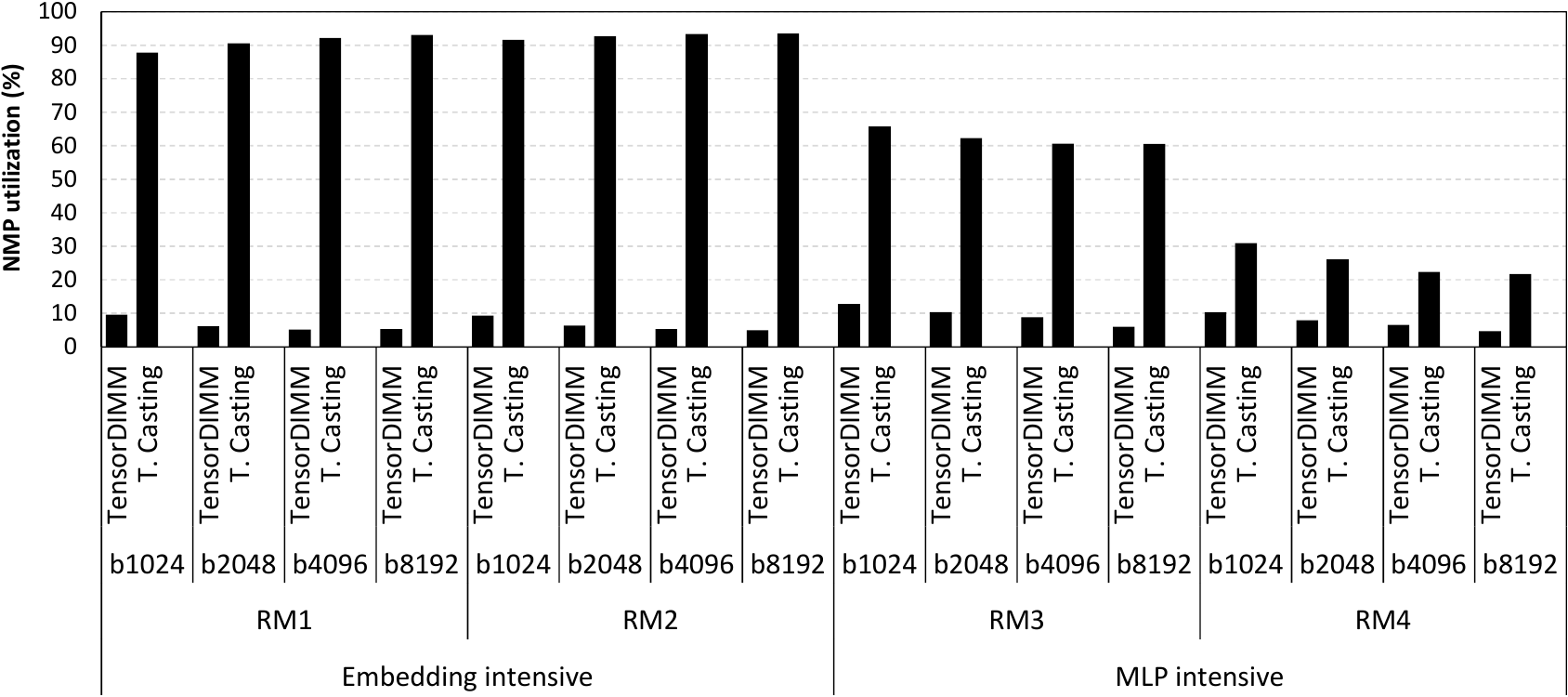}
\caption{
NMP utilization (i.e., fraction of training time when NMP is active). 
}
\vspace{-1.5em}
\label{fig:nmp_utilization}
\end{figure}

{\bf Energy-efficiency.} We utilize \texttt{powerstat} to measure CPU's socket-level
power consumption and NVIDIA's \texttt{nvidia-smi} for GPU power
measurements. To quantify the power overhead of our NMP-augmented disaggregated
memory architecture, we follow the methodology employed in \cite{tensordimm}
utilizing Micron's DDR4 system power calculator~\cite{micron_dram_power_estimator}
to quantify the 32 ranked disaggregated memory node (assuming a $128$ GB LR-DIMM DDR4
		module per each rank), which we augmented with the aforementioned Verilog HDL based power estimation of NMP units. When evaluating energy consumption, we multiply the power estimation values with each CPU, GPU, and NMP node's execution time.
\fig{fig:energy_eval} shows
the energy-efficiency of \texttt{Ours(NMP)}, where the significant training throughput improvements
directly translate into energy savings. Notice that even the software-only \texttt{Ours(CPU)} provides
noticeable energy-efficiency improvements compared to \texttt{Baseline(NMP)},
underscoring the importance of properly addressing the bottlenecks of gradient expand-coalesce.

{\bf NMP utilization.}
To evaluate how well the NMP accelerator is utilized, with/without Tensor Casting,
\fig{fig:nmp_utilization} measures the fraction of training time when NMP is
active. As depicted, the baseline NMP 
	\emph{without} Tensor Casting (i.e., TensorDIMM) is only useful in accelerating
embedding gather-reduce and gradient scatter operations (see \fig{fig:eval_latency}). As such, the NMP
core is only active
during an average $7\%$ of training time (i.e., the NMP is left idle during
		$93\%$ of the training time). Our Tensor Casting algorithm significantly improves
the utilization of the NMP because it enables all major primitives of 
both forward and backpropagation for accelerated execution. Specifically,
the NMP accelerator \emph{with} Tensor Casting is actively utilized
for
an average $92\%$ of training time for embedding intensive RM1/2 and
an average $44\%$ for MLP intensive RM3/4, a significant increase in 
NMP utility compared to
TensorDIMM (which only achieves an average $6.5\%$ and $8.5\%$ utilization for RM1/2 and RM3/4, respectively), further justifying the design overheads of NMP architectures.

\subsection{Sensitivity}
\label{sect:sensitivity}

\begin{figure}[t!] \centering
	\includegraphics[width=0.485\textwidth]{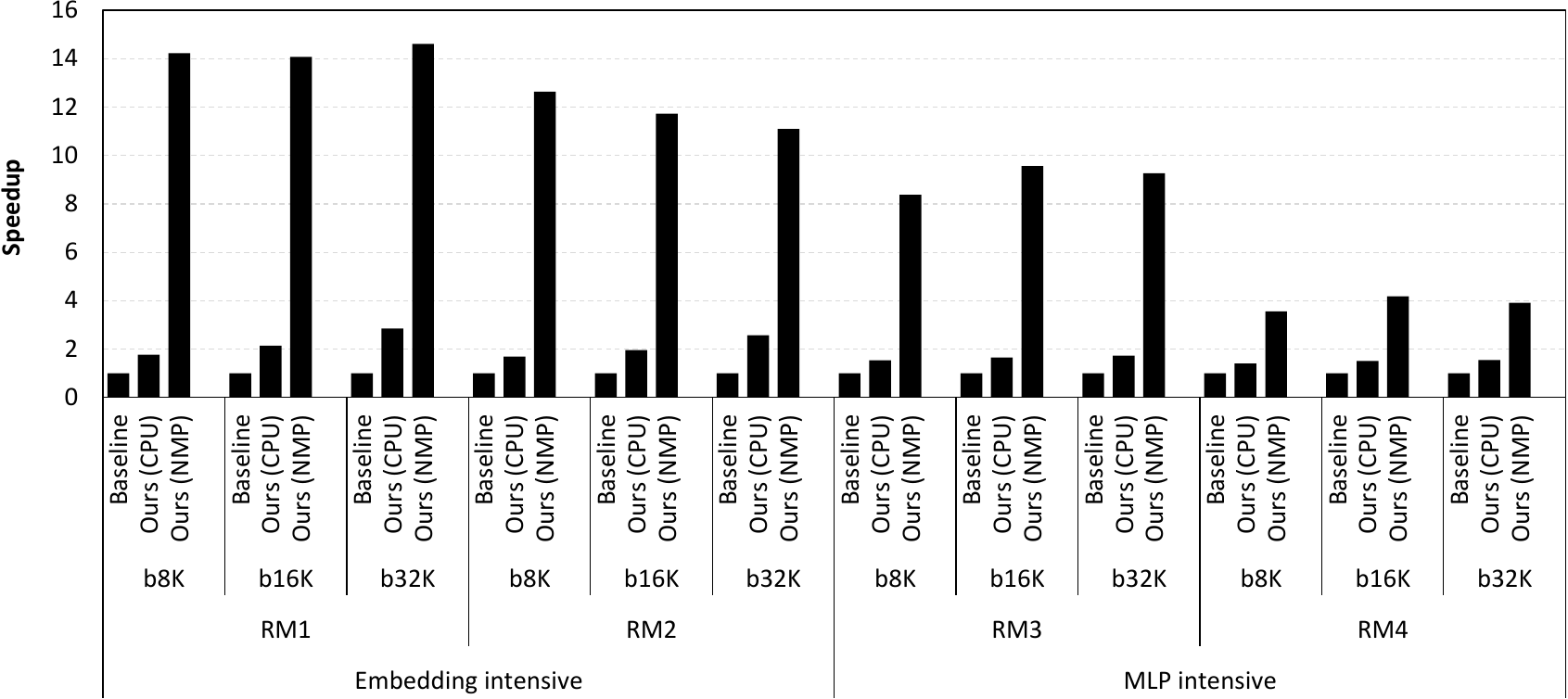}
\caption{
\tcasting sensitivity to training batch size.
}
\vspace{-1.2em}
\label{fig:eval_sensitivity_batch_sz}
\end{figure}

{\bf Training batch size.} While our nominal training batch size is chosen between
$1024$ to $4096$, several prior works employ several tens of thousands of input
batch sizes for training recommendations~\cite{mlperf:github,dlrm:github}. \fig{fig:eval_sensitivity_batch_sz} summarizes
the performance improvements \tcasting achieves with larger batch sizes, exhibiting
up to $15\times$ throughput increase. In general, we observe that the effectiveness of \tcasting remains
robust across a wide range of training batch sizes.

\begin{figure}[t!] \centering
	\includegraphics[width=0.485\textwidth]{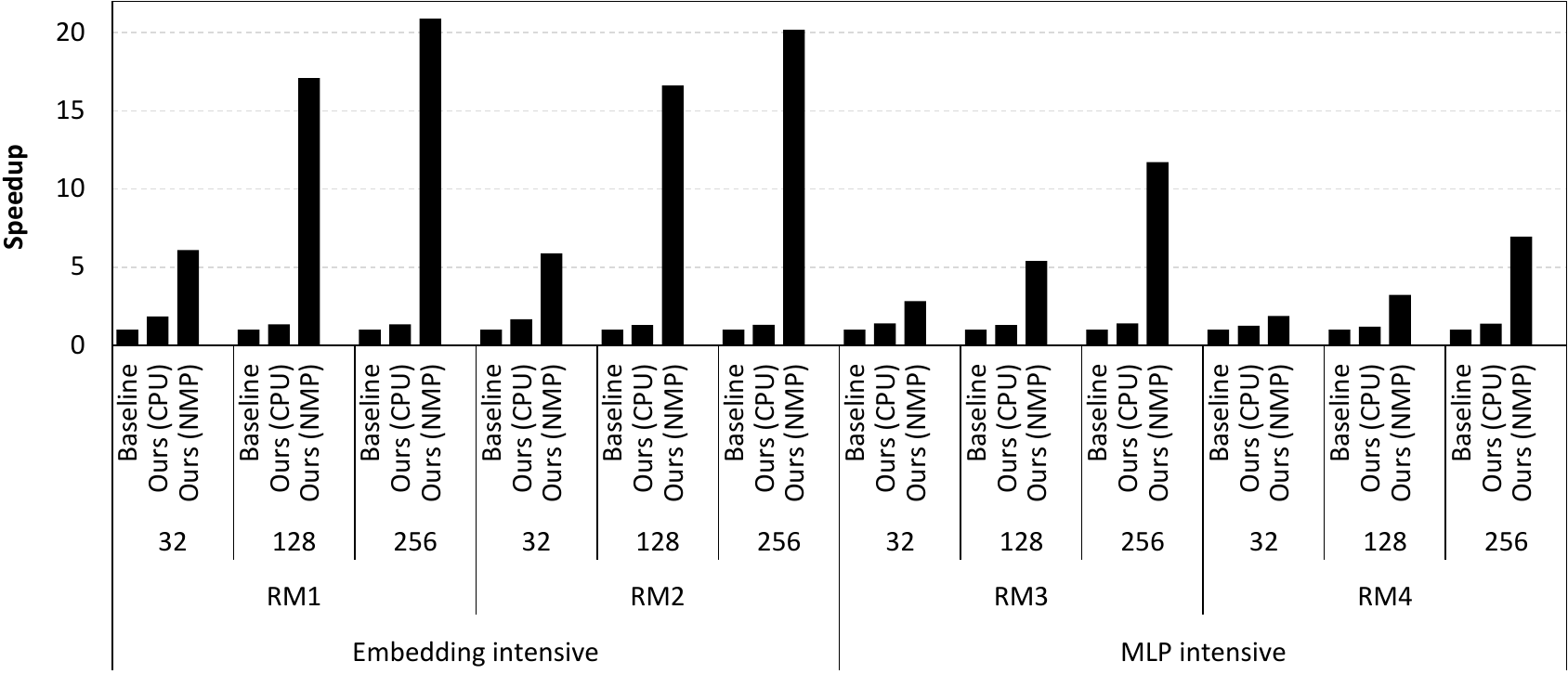}
\caption{
\tcasting sensitivity to embedding vector size.
}
\vspace{-1.2em}
\label{fig:eval_sensitivity_emb_sz}
\end{figure}

{\bf Embedding dimension size.} While our baseline embedding vector
dimension is configured as $64$, there are prior works
employing smaller~\cite{dlrm:github} or larger~\cite{youtube_recsys} embedding vector
dimensions than our assume embedding size of $64$. \fig{fig:eval_sensitivity_emb_sz} summarizes the speedup provided
with \tcasting under alternative embedding vector widths, achieving significant speedups and demonstrating
the robustness of our proposal.

{\bf Communication bandwidth.} For a conservative analysis, our default
memory-centric system assumes a modest $25$ GB/sec of communication bandwidth
between the GPU and disaggregated memory. 
	The goal of this sensitivity
study is to sweep the available communication bandwidth ($25-150$ GB/sec)
to explore how much performance is left on the table with a advanced communication
protocols such as NVLINK~\cite{nvlink}. We observe that even with our conservative baseline
with PCIe(gen3) level communication, \tcasting already achieves 
$99\%$ of the performance of a much more aggressive $150$ GB/sec configurations,
	highlighting the robustness of \tcasting. We omit the results for brevity.

\section {Conclusion}
\label{sect:conclusion}

This paper proposes \tcasting, an algorithm-architecture
co-design for training recommendation models. Unlike recent prior
literature focusing on the inference part of this emerging ML workload, the
unique contribution of our study is the exploration of this application on the
training side of things. We first provide a detailed, quantitative
analysis on training recommendation models, root-causing several system-level
bottlenecks such as gradient expand-coalesce. We then implement and demonstrate the benefits of
\tcasting on real systems, showing
that \tcasting achieves $1.9-21\times$ speedup than state-of-the-art approaches.
To the best of our knowledge, \tcasting is the first that quantitatively
explores architectural solutions tailored for training recommendation models.


\bibliographystyle{IEEEtranS}
\bibliography{ref}

\end{document}